\documentclass[prb,preprintnumbers,amsmath,amssymb,showpacs,superscriptaddress,notitlepage,nofootinbib]{revtex4-2} 

\usepackage{hyperref}
\def\doi{http://dx.doi.org/}
\usepackage{amsmath,braket,mathdots,mathtools,amssymb,xcolor,calligra,color,empheq}
\usepackage{amsthm}
\usepackage{dsfont}
\usepackage{fancyhdr}
\usepackage{color}
\usepackage{epsfig}
\usepackage{tikz}
\usetikzlibrary{spy}
\usetikzlibrary{patterns}
\usetikzlibrary{calc,arrows,positioning}
\usetikzlibrary{arrows,shadows}
\usetikzlibrary{decorations.pathmorphing}	
\usetikzlibrary{decorations.markings}
\usepackage{pgfplots}
\usepackage{pgfplotstable}
\pgfplotsset{compat=1.17} 
\usepackage{filecontents}
\usepackage[utf8]{inputenc}    

\def\sfix#1{\texorpdfstring{#1}{Lg}}

\newcommand{\be}{\begin{equation}}
\newcommand{\ee}{\end{equation}}
\newcommand{\bea}{\begin{eqnarray}}
\newcommand{\eea}{\end{eqnarray}}

\def\nn{\nonumber\\}

\def\fr#1{(\ref{#1})}
\def\ontop#1#2{\genfrac{}{}{0pt}{}{#1}{#2}}

\def\nn{\nonumber\\}

\def\fr#1{(\ref{#1})}

\def\lt{\widetilde{\lambda}}

\def\ontop#1#2{\genfrac{}{}{0pt}{}{#1}{#2}}
\newcommand{\drawbox}[1]{\vbox{\hrule\hbox{\vrule\hskip4pt\vbox{\vskip5pt
        \hbox{#1} \vskip5pt}\hskip4pt\vrule}\hrule}} 
\def\boxit#1{\drawbox{$\displaystyle #1$}}
\usepackage[most]{tcolorbox}
\tcbuselibrary{theorems}

\newtcbtheorem{Aside}{Remark}%
{breakable,colframe=blue!50!white!50!black,fonttitle=\bfseries}{}

\newtcbtheorem{Proposition}{}%
{breakable,colframe=red!50!white!50!black,fonttitle=\bfseries}{}

\begin{document}

\title{A short introduction to Generalized Hydrodynamics}
\author{Fabian H.L. Essler}
\affiliation{The Rudolf Peierls Centre for Theoretical Physics, Oxford
University, Oxford OX1 3PU, UK}

\begin{abstract}
These are notes based on lectures given at the 2021 summer school
on \emph{Fundamental Problems in Statistical Physics XV}.
Their purpose is to give a very brief introduction to
\emph{Generalized Hydrodynamics}, which provides a description of the
large scale structure of the dynamics in quantum integrable
models. The notes are not meant to be comprehensive or provide an
overview of all relevant literature, but rather give an exposition of
the key ideas for non-experts, using a simple fermionic tight-binding
model as the main example. 
\end{abstract}
\maketitle
\tableofcontents

\section{Introduction}
The notes are based on lectures given at the summer school on
\emph{Fundamental Problems in Statistical Physics XV}. 
Their purpose
is to give a very brief introduction to the exciting recent
developments in \emph{Generalized Hydrodynamics}
\cite{castro2016emergent,bertini2016transport}. They are not meant 
to be comprehensive or provide an overview of all relevant
literature. They are several recent reviews and lecture notes 
\cite{doyon2018lectures,alba2021generalized,bastianello2021introduction}
which provide much more detailed discussions and I strongly encourage the
interested reader to consult in particular the following
\begin{enumerate}
\item{} B. Doyon, \emph{Lecture notes on Generalized Hydrodynamics},
SciPost Phys. Lect. Notes p. 18 (2020).
\item{} V. Alba, B. Bertini, M. Fagotti, L. Piroli and P. Ruggeria,
  \emph{Generalized-Hydrodynamic approach to Inhomogeneous Quenches:
    Correlations, Entanglement and Quantum Effects}, 
Journal of Statistical Mechanics: Theory and Experiment 2021(11), 114004 (2021).
\end{enumerate}

\subsection{Quantum quenches and experiments}
\label{ssec:QQ}
A \emph{quantum quench} is a particular protocol for driving
many-particle quantum system out of equilibrium. It is defined as
follows.
\begin{enumerate}
\item{} The starting point is a many-particle system in a large,
finite volume $L$ with Hamiltonian $H$.
\item{} The system is then prepared in an initial state
$|\Psi(0)\rangle$, or an initial density matrix $\hat{\rho}(0)$, that has
  non-zero overlaps with exponentially many (in system size)
eigenstates of $H$. The initial state should have 
good clustering properties\footnote{By this we mean that connected
correlation functions of local operators go to zero in the limit of
large separations between them.} 
and is often taken to be a lowly entangled state.
\item{} At later times the quantum state describing the system is then
given by the solution of the time-dependent Schr\"odinger equation
\be
|\Psi(t)\rangle=e^{-iHt}|\Psi(0)\rangle.
\ee
\item{} The objective is to study expectation values of local
operators ${\cal O}_A$ in the thermodynamic limit
\be
\lim_{L\to\infty}\langle\Psi(t)|{\cal O}_A|\Psi(t)\rangle.
\ee
\end{enumerate}
Here we define \emph{local operators} as acting as the identity outside a
finite, connected spatial region in the infinite volume limit. For a
quantum spin chain operators of the form
$\sigma_{j_1}^{\alpha_1}\dots\sigma_{j_\ell}^{\alpha_\ell}$ where
$j_k\in [a,b]$ with $a,b$ fixed are local. As we will see later
locality is a very important concept in non-equilibrium dynamics. I
note that this notion of locality is very strong and a weaker notion
of \emph{quasi-locality} \cite{ilievski2016quasilocal} sufficies for
our purposes \footnote{We can loosely define quasi-local operators as
having the property that they can be approximated by a sequence of
local operators ${\cal O}_A$ defined on a finite, connected region $A$
of linear size $|A|$ such that $|\!|{\cal   O}-{\cal O}_A|\!|$ decays
faster than any power of $|A|$.} 
Often the Hamiltonian depends on a parameter $h$ such a magnetic field or
interaction strength, and a popular way of defining a quantum quench
is then to take $|\Psi(0)\rangle$ as the ground state of $H(h_0)$, and
consider time evolution under the Hamiltonian $H(h_1)$ with $h_1\neq
h_0$. This corresponds to an instantaneous ``quench'' of $h$ at time
$t=0$ from $h_0$ to $h_1$.

\begin{figure}[ht]
\begin{center}
(a)\epsfxsize=0.4\textwidth
\epsfbox{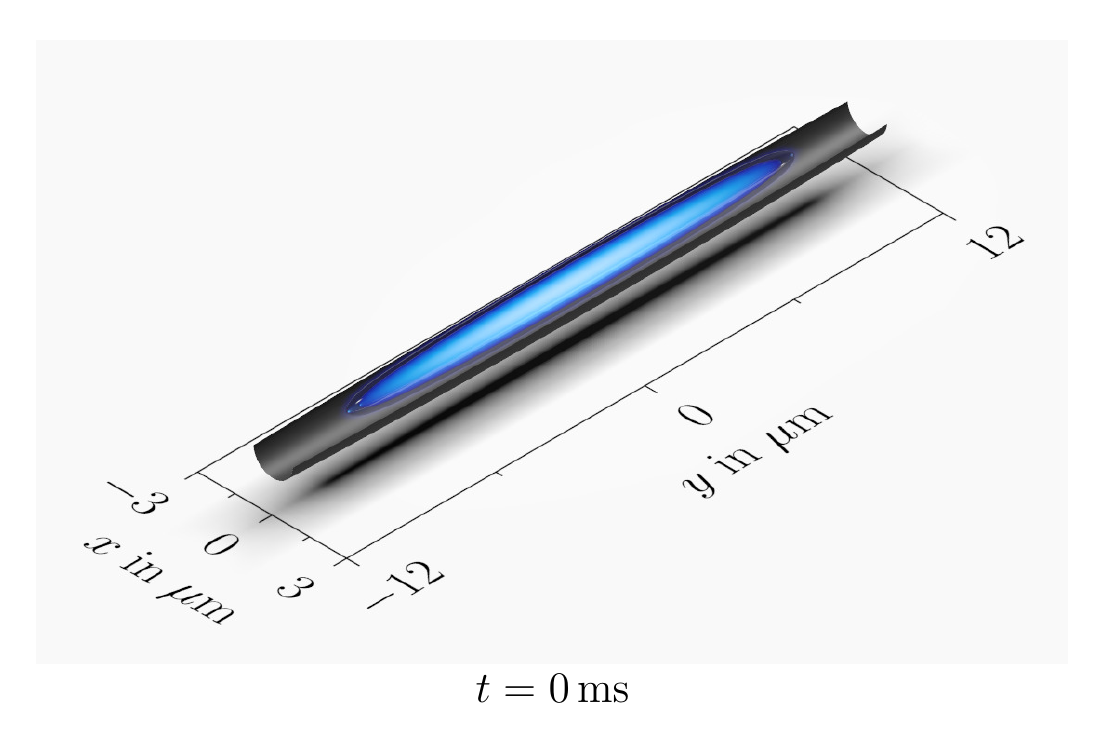}\qquad
(b)\epsfxsize=0.4\textwidth
\epsfbox{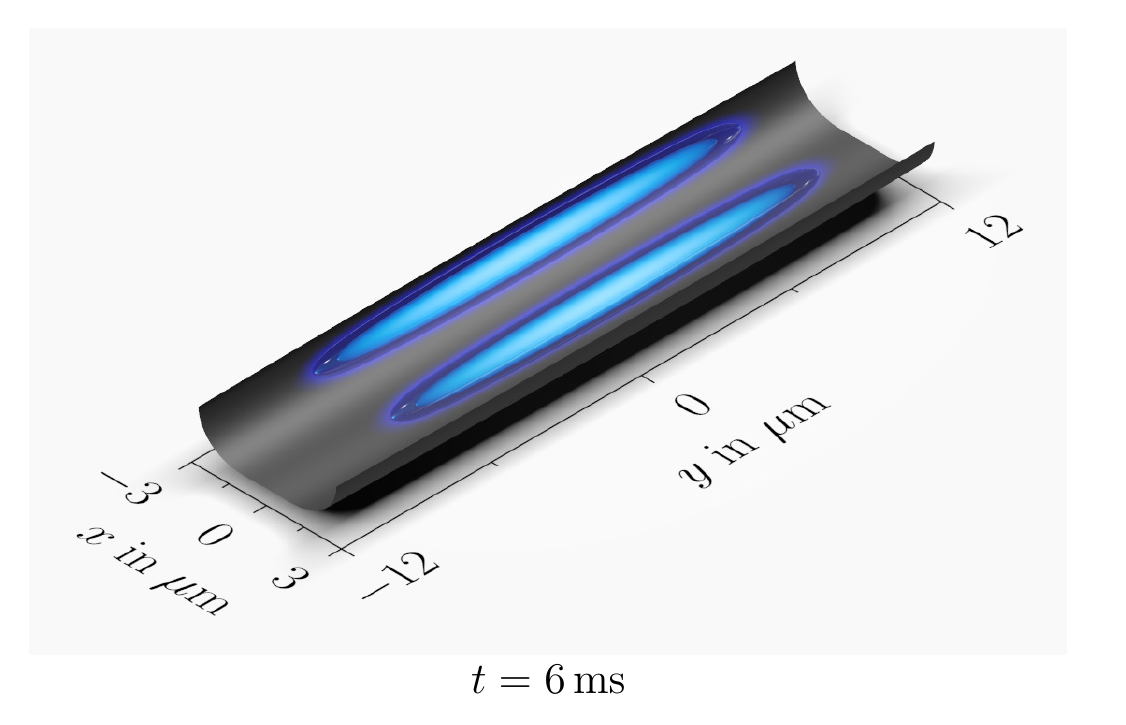}
\label{fig:split}
\end{center}
\caption{By changing the transverse confining potential a one
  dimensional Bose gas (a) is ``split into two'' (b).}
\end{figure}

The theoretical quantum quench protocol introduced above is inspired
by cold atom experiments. In order to make the connection more
concrete I now present a brief cartoon of experiments on
ultra-cold bosonic $\rm Rb$ atoms carried out e.g. in J\"org
Schmiedmayer's group in Vienna
\cite{schaff2018one,pigneur2018relaxation,vannieuwkerk2018projective,vannieuwkerk2020on,vannieuwkerk2021josephson}. The
Hamiltonian describing the atoms is to a good approximation 
\be
H(t)=\sum_j\left[-\frac{\hbar^2 \nabla_j^2}{2m}+V(\boldsymbol{r}_j,t)\right]
+\frac{g}{2}\sum_{j\neq k}\delta^{(3)}(\boldsymbol{r}_j-\boldsymbol{r}_k),
\ee
where $V(\boldsymbol{r}_j,t)$ is a confining potential that is varied
in a time-dependent way. The potential is separable in the sense that
$V(\boldsymbol{r}_j,t)=\frac{1}{2}m\omega_\parallel^2x_j^2+V_\perp(y_j,z_j,t)$ and
$V_\perp$ can be tuned in such a way that the transverse degrees of
freedom are essentially projected to the ground state(s) of the
single-particle Hamiltonian 
\be
H_{\perp,0}=\sum_j
-\frac{\hbar^2}{2m}\left[\frac{\partial^2}{\partial
    y_j^2}+\frac{\partial^2}{\partial  z_j^2} \right] +V_\perp(y_j,z_j,t).
\ee
By choosing the transverse confining potential to be very tight and
having a single minimum the system at time $t=0$ can be prepared in a
low temperature thermal state of the one dimensional Hamiltonian  
\be
H(0)\approx\sum_j\left[-\frac{\hbar^2}{2m}\frac{\partial^2}{\partial  x_j^2}
+\frac{1}{2}m\omega_\parallel^2x_j^2\right]+c\sum_{j<k}\delta(x_j-x_k)\ ,
\ee
where the interaction strength $c$ is proportional to $g$ and overlap
integrals of the eigenfunctions of the confining potential, see
e.g. Ref.~\cite{vannieuwkerk2021josephson}. 
In the experiments $V_\perp(y_j,z_j,t)$ is then changed in a
time-dependent fashion so that one ends up with a
tight double-well potential in the transverse direction, \emph{cf.}
\cite{vannieuwkerk2021josephson} for a theoretical
description. Neglecting the higher transverse modes (as they have very
high energies) then leads to a Hamiltonian that in second quantization
takes the form 
\be
H(t_0)\approx
\int dx\sum_{a}\left[\Phi^\dagger_a(x)
\Big(-\frac{1}{2m}\frac{d^2}{dx^2}+\frac{m\omega_\parallel^2}{2}x^2\Big)\Phi_a(x)
+c \Phi_a^\dagger(x)\Phi_a^\dagger(x)\Phi_a(x)\Phi_a(x)\right].
\label{Ht0}
\ee
Here $a=1,2$ label the two wells, \emph{cf.} Fig.~\ref{fig:split}, and
$\Phi_a(x)$ are canonical Bose fields obeying commutation relations
$[\Phi_a(x),\Phi^\dagger_b(y)]=\delta_{a,b}\delta(x-y)$. 
The splitting process leaves the system in some initial state
$|\Psi(t_0)\rangle$ (or more generally some initial density matrix
$\hat{\rho}(t_0)$) that is not an eigenstate of $H(t_0)$: in this 
sense the situation is analogous to a quantum quench. The system is
now left to evolve in time governed by the Hamiltonian \fr{Ht0}. At a
time $t_1$ the confining potential is switched off and the two clouds
of atoms start to expand freely in three dimensions. Eventually they
overlap and at a time $t_2$ the density of atoms is measured.
\begin{figure}[ht]
\begin{center}
\epsfxsize=0.7\textwidth
\epsfbox{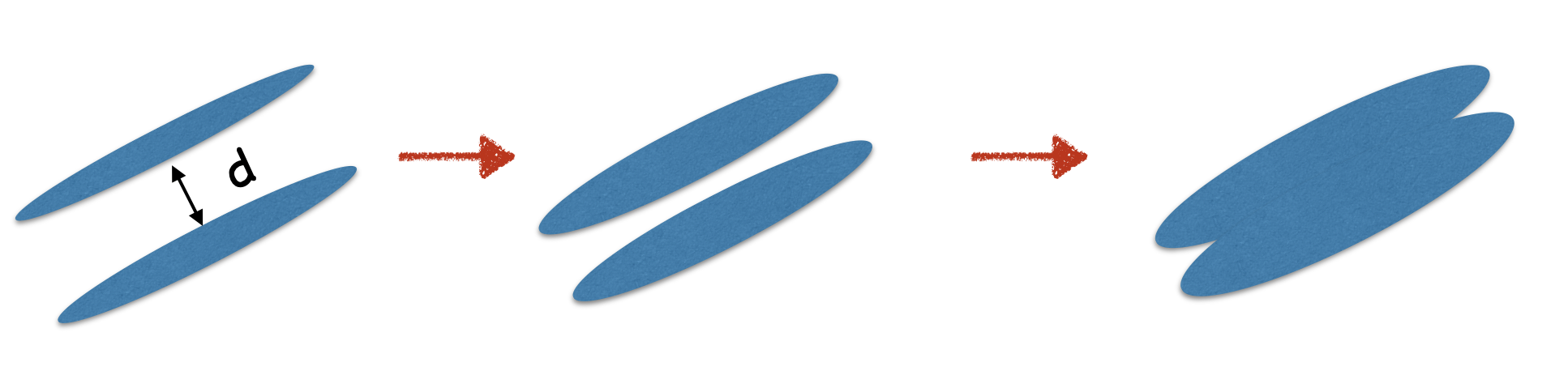}
\label{fig:expansion}
\end{center}
\caption{After switching off the confining potential the atomic clouds
  expand in three dimensions and eventually overlap.}
\end{figure}
The expansion can be easily modelled as the atoms effectively do not
interact. One therefore can integrate the Heisenberg equations of
motion for the measured observable (the particle density) backwards
and relate it to an operator in the split one dimensional Bose gas at
time $t_1$. One finds that the measured density is \cite{polkovnikov2006interference}
\be
\hat{\rho}_{\rm tof}(x,{\bf r},t_2)\approx\sum_{a,b=1}^2\int dx_1dx_2\
g_{ab}(t_2-t_1,{\bf
  r},x-x_1,x-x_2)\Phi^\dagger_a(x_1,t_1)\Phi_b(x_2,t_1) \ ,
\label{observable}
\ee
where ${\bf r}$ denote the transverse directions and $g_{ab}(t,{\bf
  r},x_1,x_2)$ are known functions. Repeating the experiment many times then
provides access to e.g. the expectation value \fr{observable} in the
split, one-dimensional Bose gas after a period of non-equilibrium
evolution.

\subsection{Homogeneous quantum quenches and local relaxation}
We first focus on the simpler case of \emph{homogeneous} systems where
both the post-quench Hamiltonian and the initial state are
translationally invariant \cite{essler2016quench}
\footnote{For lattice models we may break translational
  invariance to translations by $n$ sites, where $n$ is fixed.}.
The first question we ask is whether after a quantum quench a
many-particle system somehow relaxes, i.e. whether if we wait long
enough the quantum mechanical probability distributions describing
the outcomes of measurements become time independent. This is
equivalent to the question whether the double limit
\be
\lim_{t\to\infty}\lim_{L\to\infty}\langle\Psi(t)|{\cal O}|\Psi(t)\rangle
\ee
exists for all Hermitian operators ${\cal O}$. We note that the order of limits is
crucial here. It is easy to see that this limit cannot exist for all
observables. Indeed, let $|n\rangle$ be the eigenstates of the
Hamiltonian describing the time evolution of our system and $E_n$ the
corresponding energies. Then
\be
|\Psi(t)\rangle=\sum_n e^{-iE_nt}\langle n|\Psi(0)\rangle\ |n\rangle.
\ee
Now we can choose ``observables'' that never relax, e.g.
\be
{\cal O}={\cal O}^\dagger=|1\rangle\langle 2|+|2\rangle\langle 1|.
\ee
Indeed, we have
\be
\langle\Psi(t)|{\cal O}|\Psi(t)\rangle=A\cos\big((E_1-E_2)t+\varphi\big),
\ee
which shows that the expectation value of this particular observable
is a periodic function of time. However, the operator ${\cal O}$ is typically
highly non-local in space. This suggests that we should restrict our attention
to \emph{local measurements} and concomitantly \emph{local
  operators} ${\cal O}_A$. For these our double limit generally exists, i.e.
\be
\lim_{t\to\infty}\lim_{L\to\infty}\langle\Psi(t)|{\cal
  O}_A|\Psi(t)\rangle=\langle {\cal O}_A\rangle_{\rm stat}.
\ee
The physical picture underlying this fact is as follows. As ${\cal
  O}_A$ is a local operator it acts like the identity outside some
finite spatial region $A$. In the infinite volume limit the complement
of $A$ simply acts like a bath on $A$ and eventually leads to
relaxation. One can reformulate this observation in terms of density
matrices as follows. The density matrix of the entire system
$\hat{\rho}(t)=|\Psi(t)\rangle\langle\Psi(t)|$ in our case is a pure state
and hence can never become time independent. On the other hand, the \emph{reduced
  density matrix}
\be
\hat{\rho}_A(t)={\rm Tr}_{\bar{A}}\big[\hat{\rho}(t)\big]
\ee
describing the region $A$ on which our observable acts ($\bar{A}$ is
the complement of $A$) is a mixed state and hence can become
time-independent at late enough times in the thermodynamic limit.

A natural question to ask at this point whether it is possible to
describe the late-time limits of the expectation values of local
operators in terms of a statistical ensemble. In other words, is it
possible to find a time-independent density matrix $\hat{\rho}_{\rm SS}$
such that for any local operator ${\cal O}_A$ acting non-trivially
only on a finite subsystem $A$
\be
\lim_{t\to\infty}\lim_{L\to\infty}\langle\Psi(t)|{\cal
  O}_A|\Psi(t)\rangle=\lim_{L\to\infty}{\rm Tr}\big[\hat{\rho}_{\rm SS}{\cal O}_A\big].
\ee
We note that in analogy to equilibrium statistical mechanics (where we
can choose from micro-canonical, canonical and grand canonical ensembles)
$\hat{\rho}_{\rm SS}$ is not unique. In order to determine $\hat{\rho}_{\rm SS}$
we employ the following ergodicity principle: \emph{under time
evolution the reduced density matrix of a finite subsystem will retain
the minimal possible amount of local information on the initial state.} 
As we are dealing with an isolated quantum system with a Hamiltonian
that has a local density $H_j$ one piece of local information that is
always retained is the energy density $\langle\Psi(0)|H_j|\Psi(0)\rangle$.

\subsubsection{Local Conservation Laws}
A \emph{local conservation law} is a Hermitian operator $I^{(n)}$ that
commutes with the Hamiltonian of our system and has a density
$I^{(n)}_j$ that is a local operator as defined above, i.e.
\be
I^{(n)}=\sum_jI^{(n)}_j\ ,\quad [H,I^{(n)}]=0.
\label{LCL}
\ee
We will be particularly interested in the situation where we have many
local conservation laws that are mutually compatible, i.e.
\be
[I^{(n)},I^{(m)}]=0.
\ee
We stress that the conservation laws we have in mind here are extensive.
The existence of a local conservation law has important consequences
for the steady state density matrix $\hat{\rho}_{\rm SS}$ in translationally
invariant cases. By \fr{LCL} we
have
\be
\langle\Psi(t)|I^{(n)}|\Psi(t)\rangle=\text{time independent}.
\label{LCL2}
\ee
Translational invariance then implies that
\be
\lim_{L\to\infty}\frac{1}{L}\sum_j\langle\Psi(t)|I^{(n)}_j|\Psi(t)\rangle=
\lim_{L\to\infty}\langle\Psi(t)|I^{(n)}_j|\Psi(t)\rangle.
\label{translinv}
\ee
Combining \fr{translinv} with \fr{LCL2} we conclude that
\be
\lim_{L\to\infty}\langle\Psi(0)|I^{(n)}_j|\Psi(0)\rangle=
\lim_{t\to\infty}\lim_{L\to\infty}\langle\Psi(t)|I^{(n)}_j|\Psi(t)\rangle=
{\rm Tr}\left[\hat{\rho}_{\rm SS}\ I^{(n)}_j\right],
\label{LCL3}
\ee
where in the last step we have used that $I^{(n)}_j$ are local
operators. This tells us that \emph{$\hat{\rho}_{\rm SS}$ retains
information about the expectation values of all local conservation
laws in the initial state}.
\subsubsection{Thermalization}
As we are dealing with an isolated quantum system energy is always
conserved
\be
e_0=\lim_{L\to\infty}\frac{\langle\Psi(t)|H|\Psi(t)\rangle}{L}=
\lim_{L\to\infty}\langle\Psi(t)|H_j|\Psi(t)\rangle.
\ee
This is the minimal amount of information on the initial state
$|\Psi(0)\rangle$ that gets retained under the dynamics. If there are
no conserved quantities other than energy the system
\emph{thermalizes} at late times after a quantum quench. The steady
state density matrix is then given by a finite temperature
(equilibrium) ensemble constructed as follows. We define a Gibbs
density matrix
\be
\hat{\rho}_{\rm GE}=\frac{e^{-\beta_{\rm eff}H}}{Z_{\rm GE}},
\ee
and fix the effective temperature $\beta_{\rm eff}^{-1}$ by requiring
it to correspond to the energy density established by the choice of
initial state 
\be
e_0=\lim_{L\to\infty}\frac{{\rm Tr}\left[\hat{\rho}_{\rm GE}\ H\right]}{L}.
\ee
Under this choice we have
\be
\hat{\rho}_{\rm SS}=\hat{\rho}_{\rm GE}\ .
\ee
We could have equally well chosen a micro-canonical description
\be
\hat{\rho}_{\rm SS}=\hat{\rho}_{\rm MC}=\frac{1}{{\cal N}}\sum_{|E_n-Le|<\epsilon}|n\rangle\langle n|,
\ee
where $|n\rangle$ are energy eigenstates with energy $E_n$ and ${\cal
  N}$ is a normalization factor that ensures that ${\rm Tr}(\hat{\rho}_{\rm
  MC})=1$. Finally we note that averaging over a micro-canonical shell
is not required as long we use a typical energy eigenstate to define
our micro-canonical ensemble, which then takes the simple form
\be
\hat{\rho}_{\rm MC}=|n\rangle\langle n|\ .
\ee
Drawing an energy eigenstate at random out of our micro-canonical
window provides us with a typical state with a probability that is
exponentially close (in $L$) to one.
\subsubsection{Non-equilibrium steady states and Generalized Gibbs Ensembles}
\label{sssec:GGE}
If we have additional conservation laws with local densities
$I^{(n)}_j$ the system cannot thermalize because
\be
\lim_{L\to\infty}\langle\Psi(0)|I^{(n)}_j|\Psi(0)\rangle=
{\rm Tr}\left[\hat{\rho}_{\rm SS}\ I^{(n)}_j\right],
\label{constraints}
\ee
which tells us that the system retains more information on the initial
state than just its energy density. What should the ensemble
describing the steady state then be? The answer to this question is
provided by Rigol et al \cite{rigol07relaxation} (see also
Jaynes \cite{jaynes57information}): we should
maximize the entropy under the constraints \fr{constraints}. This
leads to a \emph{generalized Gibbs ensemble}
\be
\hat{\rho}_{\rm GGE}=\frac{1}{Z_{\rm GGE}}e^{-\sum_n\lambda_n I^{(n)}},
\ee
where the Lagrange multipliers $\lambda_n$ are fixed by
\be
i^{(n)}=\lim_{L\to\infty}\langle\Psi(0)|I^{(n)}_j|\Psi(0)\rangle=
{\rm Tr}\left[\hat{\rho}_{\rm GGE}\ I^{(n)}_j\right].
\ee
We note that solving this system of equations is a difficult task in
general. Alternatively we may employ a \emph{generalized
micro-canonical ensemble} \cite{cassidy2011generalized,caux2013time}:
\be
\hat{\rho}_{\rm GMC}=|\Phi_L\rangle\langle\Phi_L|\ ,
\ee
where $|\Phi_L\rangle$ be a simultaneous eigenstate of all local
conservation laws $I^{(n)}$ such that
\be
\lim_{L\to\infty}\left[\frac{I^{(n)}}{L}-i^{(n)}\right]|\Phi_L\rangle=0.
\ee

\subsubsection{Approach to the steady state}
\label{sssec:approach}
In integrable models how fast the expectation value of a
given operator approaches its steady state value depends on its
locality properties \emph{relative to the elementary excitations in
  the model} \cite{bertini2014quantum}. Expectation values of
operators that are local relative to the elementary excitations
approach their stationary values in a power-law fashion
\cite{essler2016quench}. The spatial ``size'' $\ell$ of the operator
under consideration\footnote{We define this as the size of the
connected spatial region on which the operator acts non-trivially.}
together with the maximal propagation velocity of elementary
excitations $v_{\rm max}$ provides a time scale $t_\ell=\ell/2v_{\rm max}$
after which relaxation to the steady state begins
\cite{calabrese2007quantum}. The rule of thumb is that the more local
an operator is, the faster its expectation value relaxes towards its
steady-state value.

\subsection{Summary}
Integrable models with short-range interactions prepared in
homogeneous initial states that have good clustering properties relax 
locally to non-thermal stationary states, which are completely specified
by the expectation values of the (quasi)local conservation laws. In
terms of the time-evolving density matrix $\hat{\rho}(t)$ of the
system this statement reads
\be
\lim_{t\to\infty}\lim_{L\to\infty}{\rm
  Tr}_{\bar{A}}\left[\hat{\rho}(t)\right]=\lim_{L\to\infty}{\rm
  Tr}_{\bar A}\left[\hat{\rho}_{\rm GGE}\right]
=\lim_{L\to\infty}{\rm
  Tr}_{\bar A}\left[\hat{\rho}_{\rm GMC}\right].
\ee
where $\bar{A}$ is the complement of an arbitrary, finite subsystem
$A$. If the following we will be interested in the situation where our
initial density matrix is inhomogeneous. The idea there is that we
still have local relaxation, but in a \emph{space and time dependent
  way}, i.e. 
\be
\lim_{t\to\infty}\lim_{L\to\infty}{\rm
  Tr}_{\bar{A}}\left[\hat{\rho}(t)\right]
=\lim_{t\to\infty}\hat{\rho}_{{\rm GMC},A}(t).
\ee

\section{Brief introduction to quantum integrable models I: free theories}

The simplest integrable models are free theories and arguably the
simplest free theory is the tight-binding chain
\be
H=\sum_{j=1}^L-J(c^\dagger_jc_{j+1}+{\rm h.c.})-\mu c^\dagger_jc_j=
\frac{1}{L}\sum_p\epsilon(p) \hat{n}(p)\ ,
\label{Hfree}
\ee
where $c_j$ and $c_j^\dagger$ are fermionic creation and annihilation
operators on site $j$ and
\be
\epsilon(p)=-2J\cos(p)-\mu\ ,\quad
\hat{n}(p)=c^\dagger(p)c(p)\ ,\quad
c(p)=\sum_{j=1}^Lc_je^{-ipj}.
\label{epsilon0}
\ee
Imposing periodic boundary conditions quantizes the momenta
\be
k_j=\frac{2\pi j}{L}\ ,\quad j=1,2,\dots,L.
\ee
The momentum space operators obey canonical anticommutation relations
of the form
\be
\{c(p),c^\dagger(q)\}=L\delta_{p,q}\ .
\ee
The $2^L$ energy eigenstates are conveniently expressed in the
momentum representation
\begin{align}
|p_1,\dots,p_N\rangle&=\frac{1}{L^{N/2}}\prod_{j=1}^Nc^\dagger(p_j)|0\rangle\ ,\quad
p_1<p_2<\dots<p_N\ ,\nn
H|p_1,\dots,p_N\rangle&=\left(\sum_{j=1}^N\epsilon(p_j)\right)|p_1,\dots,p_N\rangle.
\end{align}

\subsection{Local conservation laws and associated currents}
It follows from the representation \fr{Hfree} that all mode occupation
numbers are conserved
\be
[H,\hat{n}(k_j)]=0=[\hat{n}(k_j),\hat{n}(k_\ell)]\ .
\ee
They are in one-to-one correspondence with mutually commuting, extensive
conservation laws with local densities by \footnote{In these notations
$Q^{(0,0)}=-2J\hat{N}$, where $\hat{N}$ is the fermion number operator.}
\be
Q^{(n,\alpha)}=-(-i)^\alpha
J\sum_{j=1}^L\left[c^\dagger_jc_{j+n}+(-1)^\alpha
  c^\dagger_{j+n}c_j\right]\equiv\frac{1}{L}\sum_{j=1}^L\epsilon^{(n,\alpha)}(k_j)\ \hat{n}(k_j)\ ,\quad
\alpha=0,1.
\label{charges}
\ee
We have $H=Q^{(1,0)}$, $[Q^{(n,\alpha)},Q^{(m,\beta)}]=0$ and
\be
\epsilon^{(n,0)}(k)=-2J\cos(nk)\ ,\quad
\epsilon^{(n,1)}(k)=-2J\sin(nk)\ .
\ee
The mode occupation operators can be expressed as linear combinations
of the conservation laws as
\be
\hat{n}(k_j)=-\frac{1}{2J}\sum_{n=0}^{L-1}e^{-ik_j n}\left[Q^{(n,0)}+iQ^{(n,1)}\right].
\label{nk}
\ee
The Heisenberg equations of motion for the densities of the
conservation laws take the form of continuity equations
\begin{align}
  \frac{d}{dt}Q^{(n,\alpha)}_j&=J^{(n,\alpha)}_{j-1}-J^{(n,\alpha)}_{j}\ ,\nn
  J^{(n,\alpha)}_j&=J^2i^{1-\alpha}\left[c^\dagger_jc_{j+n+1}-c^\dagger_{j+1}c_{j+n}+(-1)^\alpha
    (c^\dagger_{j+n}c_{j+1}-c^\dagger_{j+n+1}c_{j})\right].
\label{eomQJ}
\end{align}

\subsection{Macro-states}
In the thermodynamic limit physical properties are described in terms
of \emph{macro-states} defined as follows. Consider $L$, $N$ very
large with $D=N/L$ fixed and let $0\leq \rho(k)\leq 1$ be a given
function. The family of energy eigenstates $\{|k_1,\dots,k_N\rangle\}$
such that
\be
\frac{2\pi}{L}\times\text{ number of }k_j\text{ in
}[k,k+\Delta k]=\rho(k)\Delta k
\label{macmic}
\ee
is called a \emph{macro-state}. We have
\begin{align}
\frac{1}{L}Q^{(n,\alpha)}|k_1,\dots,k_N\rangle&=
\frac{1}{L}\sum_{j=1}^N\epsilon^{(n,\alpha)}(k_j)
|k_1,\dots,k_N\rangle\ ,\nn
\frac{1}{L}\sum_{j=1}^N\epsilon^{(n,\alpha)}(k_j)&=
\int_{-\pi}^\pi\frac{dk}{2\pi}\ \rho(k)\epsilon^{(n,\alpha)}(k)+o(L^0).
\end{align}
We see that the extensive parts of the eigenvalues only depends on the
density $\rho(k)$, i.e. is the same for all micro-states. Conversely,
if we fix the expectation values of all local
conservation laws in some quantum state in the thermodynamic limit
\be
\lim_{L\to\infty}\frac{1}{L}{\rm Tr}\left[\hat\rho
  Q^{(n,\alpha)}\right]=q^{(n,\alpha)}\ ,
\ee
then relation \fr{nk} fixes a unique macro-state
\be
\rho(k)=\lim_{L\to\infty}-\frac{1}{2JL}\sum_{n=0}^{L-1}e^{-ik n}\left[q^{(n,0)}+iq^{(n,1)}\right].
\ee
\subsubsection{Counting micro-states}
There are clearly many ways of fulfilling \fr{macmic} for a given
density $\rho(k)$. The total number of possible $k_j$ values in the
interval $[k,k+\Delta k]$ is $\Delta n_{\rm
  vac}=[L\Delta k/2\pi]$, where $[x]$ denotes the integer part of
$x$. Of these ``vacancies'' $\Delta n_{\rm   p}=[\rho(k)L\Delta k/2\pi]$
are occupied. The number of ways of distributing $\Delta n_{\rm p}$
particles among $\Delta n_{\rm vac}$ vacancies is 
\be
\genfrac(){0pt}{0}{\Delta n_{\rm vac}}{\Delta n_{\rm p}}
=\frac{(\Delta n_{\rm vac})!}{(\Delta n_{\rm p})!\ (\Delta n_{\rm vac}-\Delta n_{\rm p})!}.
\label{deltaN}
\ee
This first of all shows that in general exponentially many (in $L$)
micro-states correspond to each macro-state. Using that the entropy
our macro-state is given by $S=\ln(\#\text{ of micro-states})$, the
contribution arising from reordering momenta in the interval $[k,k+\Delta k]$ is
\be
\Delta S=\ln{\genfrac(){0pt}{0}{\Delta n_{\rm vac}}{\Delta n_{\rm p}}}
\simeq
\frac{-L\Delta k}{2\pi}\left[\rho(k)\ln\big( \rho(k)\big)
+\big(1-\rho(k))\ln\big(1- \rho(k)\big)\right],
\ee
where we have used Stirling's formula in the second step. As we are
interested in very large system sizes we therefore have
\be
S[\rho]=s[\rho]L=-L\int_0^{2\pi}\frac{dk}{2\pi}\left[\rho(k)\ln\big(\rho(k)\big)
+\big(\rho_h(k))\ln\big(\rho_h(k)\big)\right]+o(L),
\ee
where we have defined the density of holes as $\rho_h(k)=1-\rho(k)$.
\subsubsection{Typical vs atypical states}
It is clear from the above construction that any positive function
$\rho(k)$ gives rise to a macro-state and these generically have
finite entropy densities in the thermodynamic limit. Importantly these
macro-states are in general not thermal. The thermal states are
obtained by extremising the free energy per site
\be
f[\rho]=\int_{-\pi}^\pi\frac{dk}{2\pi}\epsilon(k)\rho(k)-Ts[\rho]\ .
\ee
This gives
\be
\frac{\delta f[\rho]}{\delta\rho(k)}=0\Longrightarrow \rho_{\rm
  th}(k)=\frac{1}{e^{\epsilon(k)/T}+1} \ .
\ee
Thermal states are by construction the maximal entropy states,
i.e. the most likely states, at a given energy density. In free
theories other macro-states exist at the same energy density, but
their entropies are smaller than the one of the thermal state. This
means in particular that if we select a micro-state with a given
energy density at random, this state will be thermal with a
probability that is exponentially close (in system size) to one. So
typical states at a given energy density are thermal, but there are
exponentially many ``atypical'' states as well, which differ from the
thermal state by the values of the higher conservation laws and hence
have different local properties. This situation generalizes to
interacting integrable models \cite{essler2019chapter}, where such
atypical states can have very interesting properties
\cite{veness2017quantum}. 
\subsection{Expectation values of local operators in the thermodynamic limit}
\label{ssec:EVfree}
Let $|k_1,\dots,k_N\rangle$ be a micro state corresponding to a macro
state with density $\rho(k)$. Then expectation values of fermion
bilinears depend only on the macro-state up to finite-size corrections
\be
\langle k_1,\dots,k_N|c^\dagger_j
c_\ell|k_1,\dots,k_N\rangle=\frac{1}{L}\sum_{m=1}^N e^{-ik_m(j-\ell)}
=\int_0^{2\pi}\frac{dk}{2\pi}\ \rho(k)\ e^{-ik(j-\ell)}+ o(L^0).
\label{expect}
\ee
By Wick's theorem this fact gets lifted to any multi-point correlation
function involving a fixed, finite number of fermion operators. This in turn
means that expectation values of any finite number of fermion
operators calculated in different micro states corresponding to the
same macro-state differ only by finite-size corrections that vanish in
the thermodynamic limit. Given a local operator ${\cal O}_j$ we
introduce the following notations for later convenience 
\be
\langle\boldsymbol{\rho}|{\cal O}_j|\boldsymbol{\rho}\rangle=\lim_{\ontop{N,L\to\infty}{N/L=D\text{ fixed}}}
\langle k_1,\dots,k_N|{\cal
  O}_j|k_1,\dots,k_N\rangle\ .
\label{MSEV}
\ee

\subsection{``Excitations'' over (finite entropy density) macro-states}
Let us consider a micro state $|k_1,\dots,k_N\rangle$ corresponding to the macro-state
with density $\rho(k)$. We can consider ``excitations'' over
the micro state by composing elementary particle and hole excitations:
\begin{itemize}
\item{} Particle excitations
\be
c^\dagger(p)|k_1,\dots,k_N\rangle\ ,\quad p\notin\{k_j\}.
\ee
The energy and momentum of this excitation relative to those of
$|k_1,\dots,k_N\rangle$
are
\be
\Delta E_p=\epsilon(p)\ ,\qquad \Delta P_p=p\ ,
\ee
where the dispersion $\epsilon(p)$ is given by \fr{epsilon0}.
\item{} Hole excitations
\be
c(q)|k_1,\dots,k_N\rangle\ ,\quad q\in\{k_j\}.
\ee
The energy and momentum of this excitation relative to those of
$|k_1,\dots,k_N\rangle$ are
\be
\Delta E_h=-\epsilon(q)\ ,\qquad \Delta P_h=-q.
\ee
\end{itemize}
In contrast to excitations over the ground state the energy of these
``excitations'' can be either positive or negative. In the
thermodynamic limit we obtain families of excited states parametrized by
their particle and hole momenta. Importantly they propagate with a
group velocity that is simply the derivative of the ``bare''
dispersion relation 
\be
v(p)=\epsilon'(p)\ .
\ee
This is a special feature of free theories and is not the case for
interacting integrable models.

\section{Brief introduction to quantum integrable models II: interacting theories}
\label{sec:LL}
We now want to generalize the results discussed above for free
theories to interacting integrable models, i.e.
\begin{enumerate}
\item{}Identify local conservation laws $Q^{(n)}$.
\item{}Construct simultaneous eigenstates of $H$ and $Q^{(n)}$.
\item{}Construct  macro-states in the thermodynamic limit.
\item{}Work out stable excitations over these macro-states and
  determine their group velocities.
\end{enumerate}
The example we will work with is the $\delta$-function Bose gas
\cite{korepin1993quantum}, also known as the Lieb-Liniger model
\cite{lieb1963exact}
\be
H=\int dx
\ \Phi^\dagger(x)\left[-\frac{\hbar^2}{2m}\partial_x^2\right]\Phi(x)+c\int
dx \big(\Phi^\dagger(x)\big)^2\big(\Phi(x)\big)^2\ .
\ee
Here $\Phi(x)$ is a complex Bose field with canonical commutation
relations $[\Phi(x),\Phi^\dagger(y)]=\delta(x-y)$. In first
quantization the Hamiltonian for $N$ bosons reads
\be
H=\sum_{j=1}^N-\frac{\hbar^2}{2m}\frac{\partial^2}{\partial
  x_j^2}+2c\sum_{j<k}\delta(x_j-x_k)\ .
\ee
It is customary to work in notations where $\hbar=1=2m$ and we do this
from here on.
The Lieb-Liniger model is integrable and local conservation laws can be
constructed using the quantum inverse scattering method
\cite{korepin1993quantum}, and the first few read \cite{davies2011higher}
\begin{align}
Q^{(0)}&=\int dx \ \Phi^\dagger(x)\Phi(x)\ ,\qquad
Q^{(1)}=-i\int dx \ \Phi^\dagger(x)\partial_x\Phi(x)\ ,\nn
Q^{(2)}&=H\ ,\qquad Q^{(3)}= i\int dx\left[
  \Phi^\dagger(x)\partial_x^3\Phi(x)-\frac{3c}{2}\big(\Phi^\dagger(x)\big)^2\partial_x\big(\Phi(x)\big)^2\right] .
\end{align}
The Heisenberg equations of motion for the densities of these
conserved quantities take the form of continuity equations
\be
\frac{\partial}{\partial t}Q^{(n)}(x)=i[H,Q^{(n)}(x)]=
-\frac{\partial}{\partial x}J^{(n)}(x)\ .
\label{conteqint}
\ee
The first two currents are
\begin{align}
J^{(0)}&=-i\int dx \left[
    \big(\partial_x\Phi^\dagger(x)\big)\Phi(x)
    -\Phi^\dagger(x)\partial_x\Phi(x)\right] ,\nn
J^{(1)}&=-\frac{1}{2}\int dx
\left[\partial_x^2\Phi^\dagger\Phi+\Phi^\dagger\partial_x^2\Phi-
2\partial_x\Phi^\dagger\partial_x\Phi -2c
  \big(\Phi^\dagger\big)^2\big(\Phi\big)^2 \right] .
\label{currents}
\end{align}
\subsection{Simultaneous eigenstates of \sfix{$Q^{(n)}$}}
Simultaneous eigenstates of the Hamiltonian and the conservation laws
are constructed as follows. Working in the position representation
\be
|\chi\rangle=\frac{1}{\sqrt{N!}}\int dx_1\dots dx_N\
\chi(x_1,\dots,x_N)\ \Phi^\dagger(x_1)\dots\Phi^\dagger(x_N)|0\rangle
\ee
the time-independent Schr\"odinger equation for $N$-particle states reads
\be
\bigg[\sum_{j=1}^N-\frac{\partial^2}{\partial
    x_j^2}+2c\sum_{k<j}\delta(x_j-x_k)\bigg]\chi(x_1,\dots,x_N)=E\chi(x_1,\dots,x_N)\ .
\ee
When the separation between any two particles is large the solution is
simply a superposition of plane waves. The structure of the interaction
term is such that the plane-wave solutions can be consistently matched whenever
two particles meet such that the exact eigenfunctions take the form of
the celebrated \emph{Bethe ansatz}
\be
\chi_{\boldsymbol{\lambda}}(x_1,\dots
x_N)=\frac{1}{{\cal N}}\sum_{P\in S_N}{\rm
  sgn}(P)e^{i\sum_{j=1}^N\lambda_{P_j}x_j}\prod_{j>k}\big[\lambda_{P_j}-\lambda_{P_k}-ic\big]\ ,
\label{wavefn}
\ee
where the normalization is ${\cal N}^2=
N!\prod_{j>k}[(\lambda_j-\lambda_k)^2+c^2]$.
If the model were not integrable we would have a superposition of
plane waves with a set single-particle momenta $\{k_1,\dots,k_N\}$ in
the sector $x_1<x_2<\dots<x_N$, but with a different set $\{p_1,\dots,p_N\}$
in the sector  $x_2<x_1<x_3<\dots<x_N$, such that the total energy and
the total momentum are the same
\be
E=\sum_{j=1}^Nk_j^2=\sum_{j=1}^Np_j^2\ ,\quad
P=\sum_{j=1}^Nk_j=\sum_{j=1}^Np_j\ .
\ee
The presence of the higher conservation laws $Q^{(n)}$ ensures that
the set of single-particle momenta does not change between different
sectors. The states $|\chi_{\boldsymbol{\lambda}}\rangle$
corresponding to the wave functions \fr{wavefn} are in fact simultaneous
eigenstates of all conserved charges
\be
Q^{(n)}|\chi_{\boldsymbol{\lambda}}\rangle=\left(\sum_{j=1}^Nq^{(n)}(\lambda_j)\right)
|\chi_{\boldsymbol{\lambda}}\rangle\ ,\quad q_n(\lambda)=\lambda^n.
\label{Qn}
\ee
As we need a description of the system in a large, finite volume we
now impose periodic boundary conditions 
\be
\chi_{\boldsymbol{\lambda}}(x_1,\dots ,
x_j+L,x_N)=\chi_{\boldsymbol{\lambda}}(x_1,\dots x_N)\ .
\ee
As usual these lead to the quantization of the single-particle
momenta, but unlike in the free case the resulting quantization
conditions are very non-trivial
\be
e^{i\lambda_j
  L}=-\prod_{k=1}^N\frac{\lambda_j-\lambda_k+ic}{\lambda_j-\lambda_k-ic}\ ,\quad
j=1,\dots,N.
\label{BAE1}
\ee
This set of equations is referred to as \emph{Bethe equations}.
Importantly, as a result of the interactions the ``single-particle
momenta'' $k_j$ are \emph{state dependent} and correlated with one
another via the quantization conditions \fr{BAE1}. Taking the
logarithm of these equations we obtain 
\be
\lambda_jL+\sum_{k=1}^N\theta(\lambda_j-\lambda_k)
=2\pi I_j\ ,\qquad \theta(x)=2\ {\rm arctan}\big(\frac{x}{c}\big)\ ,
\label{BAE2}
\ee
where $I_j$ are integers (half-odd integers) for $N$ odd (even). The
key point is that each set $\{I_j\}$ of distinct (half-odd) integers
is in one-to-one correspondence with a solution $\{\lambda_j\}$ of the
Bethe equations, which in turn provides the wave-function
$\chi_{\boldsymbol{\lambda}}(x_1,\dots,n_N)$ of a simultaneous
eigenstate of the Hamiltonian and the conserved charges. This set of
states is complete.
\subsection{Macro-states}
We now want to construct macro-states along the same lines as for free
theories. The main complication is that the quantization conditions
\fr{BAE1}, \fr{BAE2} are non-trivial and state-dependent. The way to
deal with this complication is to work with the (half-odd) integers
$I_j$, which are independent from one another. In analogy\footnote{We
adopt the customary normalization of the density $\rho(\lambda)$ in the
interacting case, which differs from \fr{macmic} by a factor of $2\pi$.} with
\fr{macmic} we may define a density for $z_j=I_j/L$ through
\be
L\chi(z)\Delta z=\text{ number of }\frac{I_j}{L}\text{ in
}[z,z+\Delta z].
\label{macmic_int}
\ee
A positive function $\chi(z)$ specifies a macro-state, and
corresponding micro-states can the obtained by choosing sets
$\{I_j\}$ distributed according to $\chi(z)$. In practice we require
a formulation in terms of the ``particle'' distribution function of
the roots $\lambda_j$ of \fr{BAE1} defined by
\be
L\rho(\lambda)\Delta \lambda=\text{ number of }\lambda_j\text{ in
}[\lambda,\lambda+\Delta \lambda].
\label{macmic_rap}
\ee
This can be related to $\chi(z)$ by turning the sum over roots in
\fr{BAE2} into an integral over the particle density $\rho(\lambda)$
in the thermodynamic limit 
\be
z_j=\frac{I_j}{L}=\frac{\lambda_j}{2\pi}+\frac{1}{2\pi
  L}\sum_{k=1}^N
\theta(\lambda_j-\lambda_k)
  \simeq\frac{\lambda_j}{2\pi}+\frac{1}{2\pi }\int_{-\infty}^\infty
  d\mu\ \theta(\lambda_j-\mu)
  \ \rho(\mu)\ .
\ee
In the thermodynamic limit we have
\be
z(\lambda)=\frac{\lambda}{2\pi}+\frac{1}{2\pi }\int_{-\infty}^\infty
d\mu\ \theta(\lambda-\mu)
\ \rho(\mu)\ .
\label{count}
\ee
The function $z(\lambda)$ is called \emph{counting function}.
It is easy to see that $z(\lambda)$ is a strictly monotonically
increasing function. We are now in a position to relate $\chi(z)$ to
$\rho(\lambda)$ by equating the numbers of roots and integers in
corresponding intervals 
\be
\rho(\lambda)d\lambda=\underbrace{\chi(z(\lambda))}_{\vartheta(\lambda)}\frac{dz}{d\lambda}d\lambda\ .
\label{chirho}
\ee
The function $\vartheta(\lambda)$ is known as \emph{occupation function}.
This gives
\be
\frac{1}{2\pi}+\int\frac{d\mu}{2\pi}
K(\lambda-\mu)\ \rho(\mu)
=\frac{\rho(\lambda)}{\vartheta(\lambda)}\ ,\qquad
K(\lambda)=\frac{2c}{c^2+\lambda^2}.
\label{TLBAE}
\ee
It is customary to define a \emph{hole density} $\rho_h(\lambda)$ by
\be
\frac{\rho(\lambda)}{\vartheta(\lambda)}=\rho(\lambda)+\rho_h(\lambda).
\label{occfn}
\ee
This establishes that rather than specifying a macro-state by the
corresponding function $\chi(z)$, we can specify it through its root
density $\rho(\lambda)$.
\subsubsection{Expectation values of conserved charges in the
  thermodynamic limit}
\label{ssec:EVmacro}
Let $|\lambda_1,\dots,\lambda_N\rangle$ be a micro state 
corresponding to a macro-state with root density $\rho(\lambda)$ 
\footnote{More
precisely we consider $N,L\gg 1$, $N/L=D$ and impose that
$\{\lambda_1,\dots,\lambda_N\}$ solve the Bethe equations
\fr{BAE2}.}. By translational invariance the densities $Q^{(n)}(x)$ of
the conserved charges fulfil
\be
\langle \lambda_1,\dots,\lambda_N|Q^{(n)}(x)|\lambda_1,\dots,\lambda_N\rangle
=\frac{1}{L}\sum_{j=1}^Nq^{(n)}(\lambda_j)=\int_{-\infty}^\infty d\lambda\
q_n(\lambda)\ \rho(\lambda) + o(L^0).
\label{LLexpect}
\ee
In the thermodynamic limit these expectation values do not depend on
the choice of micro-state. For later convenience we introduce the
notation
\be
\langle\boldsymbol{\rho}|{\cal O}(x)|\boldsymbol{\rho}\rangle=\lim_{\ontop{N,L\to\infty}{N/L=D\text{ fixed}}}
\langle \lambda_1,\dots,\lambda_N|{\cal
  O}(x)|\lambda_1,\dots,\lambda_N\rangle\ .
\ee
For local operators these expectation values are independent of the
sequence of micro-states chosen in the limiting procedure and depend
only on the root density $\rho(\lambda)$ that specifies the
macro-state \cite{korepin1993quantum,granet2020a,granet2021systematic}. 
\subsection{Stable excitations over macro-states}
\label{ssec:stable}
\begin{figure}[ht]
\begin{center}
\begin{tikzpicture}[scale=1]
\draw[->]     (3.5,0.25) arc (180: 0:1.5);
\node at (-1,0) {.};
\filldraw[black] (-0.5,0) circle (3pt);
\node at (0,0) {.};
\filldraw[black] (0.5,0) circle (3pt);
\node at (1,0) {.};
\filldraw[black] (1.5,0) circle (3pt);
\filldraw[black] (2,0) circle (3pt);
\node at (2.5,0) {.};
\node at (3,0) {.};
\draw[red] (3.5,0) circle (3pt)
node[below=5pt,black] {\small $I_h$};
\filldraw[black] (4,0) circle (3pt);
\filldraw[black] (4.5,0) circle (3pt);
\filldraw[black] (5,0) circle (3pt);
\node at (5.5,0) {.};
\filldraw[black] (6,0) circle (3pt);
\filldraw[red] (6.5,0) circle (3pt)
node[below=5pt,black] {\small $I_p$};
\node at (7,0) {.};
\filldraw[black] (7.5,0) circle (3pt);
\node at (8,0) {.};
\node at (8.5,0) {.};
\filldraw[black] (9,0) circle (3pt);
\node at (9.5,0) {.};
\filldraw[black] (10,0) circle (3pt);
\filldraw[black] (10.5,0) circle (3pt);
\filldraw[black] (11,0) circle (3pt);
\node at (11.5,0) {.};
\filldraw[black] (12,0) circle (3pt);
\node at (12.5,0) {.};
\end{tikzpicture}
\end{center}
\caption{Set of (half-odd) integers $\tilde{I}_j$ corresponding to a
particle-hole excitation over a micro state characterized by $\{I_j\}$:
one (half-odd) integer is changed from $I_h$ to $I_p$.}
\label{fig:ph}
\end{figure}
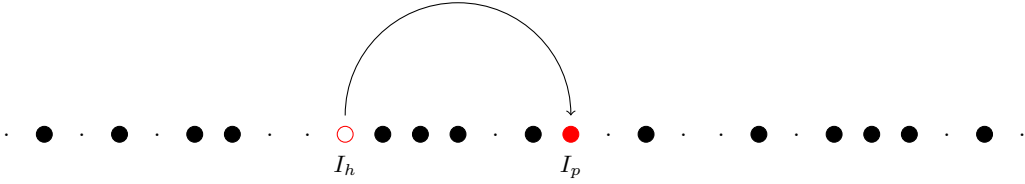

We now turn to the construction of excitations over macro-states. Our
discussion follows the textbook \cite{korepin1993quantum}.
We start by specifying a macro-state through its root density
$\rho(\lambda)$. We then determine the counting function from
\fr{count}, invert it, and determine $\chi(z)$ from \fr{chirho}
\be
\chi(z)=\rho(\lambda(z))\frac{d\lambda}{dz}.
\ee
Finally we use $\chi(z)$ to generate a distribution of $I_j/L$ that
corresponds to our given macro-state \footnote{The corresponding
set of roots $\{\lambda_j\}$ is obtained by numerically solving the
logarithmic form \fr{BAE2} of the Bethe equations, which is
straightforward to do (and is much more stable than numerically
solving \fr{BAE1}).}. We can then construct families of ``excited
states'' as we now explain for the particular example of a
``particle-hole'' excitation. The latter is obtained by changing one
of the (half-odd) integers $I_j$ as shown in Fig.~\ref{fig:ph}. 
Allowing $I_h$ and $I_p$ to vary gives rise to a two-parameter family
of eigenstates. The Bethe equations for the eigenstates characterized
by $\{I_j\}$ and $\{\tilde{I}_j\}$ are respectively
\begin{align}
\lambda_jL+\sum_{k=1}^N\theta(\lambda_j-\lambda_k)
&=2\pi I_j\ ,\nn
\lt_jL+\sum_{k=1}^N\theta(\lt_j-\lt_k)
&=2\pi \tilde{I}_j\ .
\end{align}
The two equations involving $I_p$ and $I_h$ simply fix the positions
of the corresponding momenta $\lambda_p$ and $\lambda_h$. Turning sums
into integrals and using that both micro states correspond to the same
macro-state, i.e. are described by the same root distribution
$\rho(\lambda)$, we have 
\be
\lambda_a+\int_{-\infty}^\infty d\mu
\ \theta(\lambda_a-\mu)\rho(\mu)=\frac{2\pi I_a}{L}+{\cal
  O}(L^{-1})\ ,\quad a=p,h.
\ee
Focusing on the subset of $N-2$ equations for which $I_j=\tilde{I_j}$
and taking differences we have
\be
\lambda_j-\lt_j+\frac{1}{L}\sum_{k=1}^N\theta(\lambda_j-\lambda_k)
-\theta(\lt_j-\lt_k)=0.
\ee
Next we use that $\lambda_j-\lt_j={\cal O}(L^{-1})$ to Taylor-expand,
which gives
\begin{align}
&L(\lambda_j-\lt_j)\left[1+\frac{1}{L}\sum_k K(\lambda_j-\lambda_k)\right]
-\sum_{k=1}^NK(\lambda_j-\lambda_k)(\lambda_k-\lt_k)
=\theta(\lambda_j-\lambda_p)-\theta(\lambda_j-\lambda_h).
\label{SF0}
\end{align}
Defining a \emph{shift function} by
\be
F(\lambda_j)=\frac{\lambda_j-\lt_j}{\lambda_{j+1}-\lambda_j}\ ,
\ee
we can turn \fr{SF0} into an integral equation in the thermodynamic
limit
\be
F(\lambda|\lambda_p,\lambda_h)-\int_{-\infty}^\infty
\frac{d\mu}{2\pi}\ \vartheta(\lambda)K(\lambda-\mu)\ F(\mu|\lambda_p,\lambda_h)= 
\frac{\vartheta(\lambda)}{2\pi}\left[\theta(\lambda-\lambda_p)-
\theta(\lambda-\lambda_h)\right].
\label{SF1}
\ee
We can write this as 
\be
F(\lambda|\lambda_p,\lambda_h)=f(\lambda,\lambda_p)-f(\lambda,\lambda_h)\ ,
\ee
where
\be
f(\lambda|\lambda')-\int_{-\infty}^\infty
\frac{d\mu}{2\pi}\ \vartheta(\lambda)K(\lambda-\mu)\ f(\mu|\lambda')= 
\frac{\vartheta(\lambda)}{2\pi}\theta(\lambda-\lambda').
\label{SF1a}
\ee
Using the shift function we can work out the difference in eigenvalues
of any of the conserved charges \fr{Qn}
\begin{align}
\Delta Q^{(n)}&\equiv \sum_{j=1}^N\bigg(q^{(n)}(\lt_j)-q^{(n)}(\lambda_j)\bigg)
\approx q^{(n)}(\lambda_p)-q^{(n)}(\lambda_h)+\sum_{j=1}^{N-2}
q^{(n)'}(\lambda_j)(\lt_j-\lambda_j)\nn
&=
\lambda_p^n-\lambda_h^n-\int_{-\infty}^\infty
d\mu\ F(\mu)\ n\mu^{n-1}+{\cal O}(L^{-1}).
\label{deltaqn}
\end{align}
In particular $n=1,2$ correspond to the excitation momentum and energy
respectively. We stress that the latter can be positive or
negative. We see that $\Delta Q^{(n)}$ can be expressed in the form
\begin{empheq}
[box=\fbox]{align}
\label{dQ}
\Delta Q^{(n)}&=q_n(\lambda_p)-q_n(\lambda_h)\ ,\nn
q_n(\lambda)&=\lambda^n-  \int_{-\infty}^\infty
\frac{d\mu}{2\pi}n\mu^{n-1}\ f(\mu|\lambda)\ .
\end{empheq}
Crucially, we observe that the contributions of the particle and the
hole excitation are \emph{additive}. It also should be clear that the
above construction straightforwardly generalizes to excitations
involving multiple particles and holes. Specifying $n=1,2$ in \fr{dQ}
tells us that the energy and momentum of a particle with rapidity
$\lambda_p$ are given by $q_2(\lambda_p)$ and $q_1(\lambda_p)$
respectively. The corresponding group velocity is then
\be
\boxit{v_{\boldsymbol{\rho}}(\lambda_p)=\frac{q_2'(\lambda_p)}{q_1'(\lambda_p)}.}
\label{vrho}
\ee
Here we have introduced the index $v_{\boldsymbol{\rho}}$ to indicate
that, in contrast to the case of non-interacting models, the group
velocity now depends on the macro-state under
consideration. Inspection of \fr{vrho} and \fr{dQ} shows that the
group velocity in fact depends on the macro-state only through the
occupation function $\vartheta(\lambda)$ \fr{chirho}.

\section{Hydrodynamic description of non-equilibrium dynamics}
In order to see how the non-equilibrium dynamics of many-particle
quantum systems relates to classical hydrodynamics we follow
Ref.\cite{doyon2018lectures}. 

Classical hydrodynamics in 1+1 dimensions describes the dynamics of
fluids on intermediate time and length scales. It combines the
continuity equation expressing mass conservation with the Euler equation
\begin{align}
\partial_t\rho(x,t)+\partial_x\left[v(x,t)\ \rho(x,t)\right]&=0\ ,\nn
\partial_tv(x,t)+v(x,t)\partial_xv(x,t)&=-\frac{1}{\rho(x,t)}\partial_xP[\rho(x,t)]\ .
\label{CH}
\end{align}
Here $\rho(x,t)$ is the density of the fluid, $v(x,t)$ the velocity
field and the pressure $P$ is assumed to be a function of the density
only. Classical hydrodynamics can be reformulated in terms of
continuity equations for conserved quantities by introducing the
momentum $p(x,t)=v(x,t)\rho(x,t)$ and its associated current
$j(x,t)=P+v^2(x,t)\rho(x,t)$
\begin{align}
\partial_t\rho(x,t)+\partial_xp(x,t)&=0\ ,\nn
\partial_tp(x,t)+\partial_xj(x,t)&=0\ .
\label{CH2}
\end{align}
In these lectures we want to discuss how to obtain similar
descriptions for many-particle quantum systems. The general idea is as
follows. Let $\hat\rho(0)$ be the initial density operator of a
many-particle system with (local) time-independent Hamiltonian $H$. Then the
time evolution operator is
\be
U(t)=e^{-iHt}\ ,
\ee
and the Schr\"odinger equation implies that the density matrix at time
$t$ is
\be
\hat{\rho}(t)=U(t)\hat\rho(0)U^\dagger(t)\ .
\label{timeev}
\ee
The quantities of interest are e.g. expectation values of local operators
\be
   {\rm Tr}\left[\hat\rho(t)\hat{O}(x)\right]=
   {\rm Tr}\left[\hat\rho(0)\hat{O}(x,t)\right]\ ,
   \ee
   where we have defined the Heisenberg picture operator
\be
\hat{O}(x,t)=U^\dagger(t)\hat{O}(x) U(t)\ .
\ee
As the time evolution \fr{timeev} is unitary energy is always
conserved. Let us assume that in addition to energy we have other
local conservation laws $[Q^{(n)},H]=0=[Q^{(n)},Q^{(m)}]$
\be
Q^{(n)}=\int dx\ Q^{(n)}(x)\ ,\quad Q^{(n)}(x) \text{ a local operator}.
\ee
The Heisenberg equations of motion for the density of the conservation
laws take the form of continuity equations
\be
\partial_t Q^{(n)}(x,t)=i[H,Q^{(n)}(x,t)]=-\partial_x J^{(n)}(x,t)\ ,
\ee
where $J^{(n)}(x)$ are currents associated with the conserved
quantities $Q^{(n)}$. Let us now consider the expectation values in
some quantum mechanical state
\begin{align}
q_n(x,t)&={\rm Tr}\left[\hat\rho(t) Q^{(n)}(x)\right]\ ,\nn
j_n(x,t)&={\rm Tr}\left[\hat\rho(t) J^{(n)}(x)\right]\ .
\end{align}
The basic idea is that we now choose to look at our system at length
and times scales that are large compared to some mesoscopic ``Euler''
scale. We will then assume that our system has \emph{relaxed locally}
to a state that by our ``minimal information principle'' is fully
characterized by the expectation values of the conserved quantities,
i.e.
the $q_n(x,t)$. This implies in particular that
\be
j_n(x,t)=j_n[\{q_m(x,t)\}]\ ,
\ee
and hence
\be
\partial_x j_n(x,t)=\frac{\partial j_n}{\partial q_m}\frac{\partial
  q_m}{\partial x}\equiv \sum_m
A_{n,m}[\{q_m(x,t)\}]\ \partial_xq_m(x,t)\ .
\ee
Assuming that the matrix $A$ is diagonalizable
\be
RAR^{-1}=\text{diag}(v_1,v_2,\dots)\ ,\quad
R=R[\{q_m(x,t)\}]\ ,\quad
v_\alpha=v_\alpha[\{q_m(x,t)\}]\ ,
\ee
and changing variables (locally) from $q_n(x,t)$ to $\vartheta_n(x,t)$
defined by $R_{n,m}=\frac{\partial q_n}{\partial \vartheta_m}$ we arrive at
evolution equations for the ``hydrodynamic normal modes''
\be
\boxit{\partial_t\vartheta_n(x,t)+v_n(x,t)\partial_x\vartheta_n(x,t)=0\ .}
\ee
The difficulty in this description is hidden in the fact that the
normal model velocities $v_n(x,t)$ depend on the conservation laws in
an unknown way. As we will see, it is however possible to work them
out exactly in the case of integrable models. This then allows one to
determine $\vartheta_n(x,t)$, which by our assumption of local
relaxation/minimal information completely fixes the density operator
and thus expectation values of local operators!

\subsection{``Derivation'' of GHD in free theories using a local
  density approximation} 
In a homogeneous macro-state with density $\rho(k)$ we have
(using translational invariance) in the notations of \fr{MSEV}
\begin{align}
\langle\boldsymbol{ \rho}| Q_j^{(n,\alpha)}|\boldsymbol{ \rho}\rangle
&=\int\frac{dk}{2\pi}\epsilon^{(n,\alpha)}(k) \rho(k)\ ,\nn
\langle\boldsymbol{ \rho}| J_j^{(n,\alpha)}|\boldsymbol{ \rho}\rangle
&=\int\frac{dk}{2\pi}\epsilon'(k)\ \epsilon^{(n,\alpha)}(k) \rho(k)\ .
\end{align}
Here we have used \fr{charges} and \fr{eomQJ}. Now imagine that we
prepare our system in an equilibrium density matrix $\hat\rho$ that is
not homogeneous but varies very slowly in space. Locally this density
matrix will ``look like'' a macro-state with a position-dependent
density $\rho_{x,0}(k)$, $x=ja_0$ with $a_0$ the lattice spacing, and in
particular  
\begin{align}
{\rm Tr}\left[\hat\rho Q_j^{(n,\alpha)}\right]
&\approx\int\frac{dk}{2\pi}\epsilon^{(n,\alpha)}(k) \rho_{x,0}(k)\ ,\nn
{\rm Tr}\left[\hat\rho J_j^{(n,\alpha)}\right]
&\approx\int\frac{dk}{2\pi}\epsilon'(k)\ \epsilon^{(n,\alpha)}(k) \rho_{x,0}(k)\ .
\end{align}
Let us now turn to the situation at sufficiently late times after our
quantum quench. We expect that our system will have relaxed locally,
and as long as the system varies very slowly in space we again can
employ a description in terms of an appropriately chosen macro-state, which
however now will depend on $x$ and $t$
\begin{align}
{\rm Tr}\left[\hat\rho Q_j^{(n,\alpha)}(t)\right]&=
{\rm Tr}\left[\hat\rho(t) Q_j^{(n,\alpha)}\right]
\approx\int\frac{dk}{2\pi}\epsilon^{(n,\alpha)}(k) \rho_{x,t}(k)\ ,\nn
{\rm Tr}\left[\hat\rho J_j^{(n,\alpha)}(t)\right]&=
{\rm Tr}\left[\hat\rho(t) J_j^{(n,\alpha)}\right]
\approx\int\frac{dk}{2\pi}\epsilon'(k)\ \epsilon^{(n,\alpha)}(k) \rho_{x,t}(k)\ .
\label{QJ}
\end{align}
Using the equations of motion \fr{eomQJ} for the charge densities in
\fr{QJ} we obtain an evolution equation for $\rho_{x,t}(k)$ 
\be
\int\frac{dk}{2\pi}\epsilon^{(n,\alpha)}(k)\left[
\partial_t\rho_{x,t}(k)+\epsilon'(k)\partial_x\rho_{x,t}(k)  \right]=0\ .
\ee
As the $\epsilon^{(n,\alpha)}(k)$ form a basis of periodic functions
on the interval $[-\pi,\pi]$ the term in brackets must vanish, i.e.
\be
\boxit{\partial_t\rho_{x,t}(k)+\epsilon'(k)\partial_x\rho_{x,t}(k)=0\ .}
\label{GHDFF}
\ee
These are indeed the GHD equations for free fermions. The argument
above is based on the principle of local relaxation. We stress that
there is no finite time scale over which integrable models relax
locally, instead \fr{QJ} become exact only in the so-called \emph{Euler
scaling limit} $x,t\to\infty$, $x/t$ fixed. At finite times there are
power-law corrections to \fr{QJ} and at sufficiently short times
\fr{QJ} generically do not hold.

\subsection{Dynamics for weakly inhomogeneous initial states in free theories}
The above ``derivation'' was rather hand waiving, and we now want to
do a more proper job.
The solution of the Heisenberg equation of motion for the fermion
annihilation operator in momentum space is
\be
c(p,t)=e^{-i\epsilon(p)t}c(p)\ .
\ee
Using this we can calculate the two-point functions for an arbitrary
initial density matrix $\hat{\rho}(0)$
\begin{align}
G(j,\ell;t)&={\rm Tr}\left[\hat\rho(t) c^\dagger_jc_\ell\right]=\frac{1}{L^2}\sum_{p,q}
{\rm Tr}\left[\hat\rho(0) c^\dagger(p)c(q)\right]
e^{it[\epsilon(p)-\epsilon(q)]-ijp+i\ell q}\ ,\nn
F(j,\ell;t)&={\rm Tr}\left[\hat\rho(t) c_jc_\ell\right]=\frac{1}{L^2}\sum_{p,q}
{\rm Tr}\left[\hat\rho(0) c(p)c(q)\right]
e^{-it[\epsilon(p)+\epsilon(q)]+ijp+i\ell q}\ .
\label{GFs}
\end{align}
Assuming that $\hat{\rho}(0)$ is Gaussian all higher point correlators
can then simply be obtained using Wick's theorem. This is the full
description of the dynamics. In order to analyze weakly inhomogeneous
initial states we introduce a Wigner function (for $L\to\infty$) by
\begin{align}
W(z,p;t)&={\rm Tr}\left[\hat\rho(0) \hat{W}(z,p;t)\right]\ ,\nn
\hat{W}(z,p;t)&=\int \frac{dq}{2\pi}c^\dagger\big(p-\frac{q}{2}\big)
  c\big(p+\frac{q}{2}\big)e^{iqz+it[\epsilon(p-\frac{q}{2})-\epsilon(p+\frac{q}{2})]}\ ,\quad 2z\in\mathds{Z}.
\end{align}
The Wigner function is defined for half-integer values of $z$ and has
the following properties: 
\begin{itemize}
\item[\lbrack W1.\rbrack] The single-particle Green's function can be recovered from the
  Wigner function using
  \be
{\rm Tr}\left[\hat\rho(t)
  c^\dagger(p_1)c(p_2)\right]=\frac{1}{2}\sum_{2z\in\mathds{Z}} 
e^{iz(p_1-p_2)}W(z,\frac{p_1+p_2}{2};t)\ .
\ee
\item[\lbrack W2.\rbrack] 
For homogeneous states $\hat\rho(0)$ the
Wigner function reduces to the density of states in momentum space
\begin{align}
W(z,p;t)&=\int \frac{dq}{2\pi}e^{iqz+it[\epsilon(p-\frac{q}{2})-\epsilon(p+\frac{q}{2})]}
\underbrace{{\rm Tr}\left[\hat\rho(0)c^\dagger\big(p-\frac{q}{2}\big)
    c\big(p+\frac{q}{2}\big)\right]}_{2\pi\delta(q)\rho(p)}\nn
&=\rho(p)=\sum_j\ {\rm Tr}\left[\hat\rho(0)c^\dagger_{j+1}c_1\right]e^{ipj}.
\end{align}
\item[\lbrack W3.\rbrack] 
For states $\hat\rho(t)$ that are slowly varying in space and have
short-ranged correlations the Wigner function can be interpreted as a
position-dependent density of states in momentum space. To see this we
note that the Wigner function can be expressed as
\begin{align}
W(z,p;t)&=\sum_{j,\ell}{\rm Tr}\left[\hat{\rho}(t)\ c^\dagger_jc_\ell\right]
e^{ip(j-\ell)}\ \frac{\sin\big(\pi (z-\frac{j+k}{2})\big)}{\pi \big(z-\frac{j+k}{2}\big).}
\end{align}
Let $A$ be a region centered around $z$ such that its linear size
$|A|$ is large compared to the correlation length $\xi(z)$ of the state
$\hat{\rho}(t)$, but small compared to the scale $\zeta(z)$ on which
$\hat{\rho}(t)$ becomes inhomogeneous in space. Let's further define
an integer $m(z)$ such that $\xi(z)\ll m(z)\ll \zeta(z)$. All length scales
depend on the position $z$. Then we have
\begin{align}
W(z,p;t)&=\sum_n\sum_{m}{\rm
  Tr}\left[\hat{\rho}(t)\ c^\dagger_{n+m}c_n\right]e^{ipm}
\frac{\sin\big(\pi (z-n-\frac{m}{2})\big)}{\pi
  \big(z-n-\frac{m}{2}\big)}\nn
&\approx
\sum_{m=-m(z)}^{m(z)}{\rm
  Tr}\left[\hat{\rho}(t)\ c^\dagger_{[z]+m}c_{[z]}\right]e^{ipm}
\sum_{n\in A}\frac{\sin\big(\pi (z-n-\frac{m}{2})\big)}{\pi
  \big(z-n-\frac{m}{2}\big)}\nn
&\approx
\sum_{m=-m(z)}^{m(z)}{\rm
  Tr}\left[\hat{\rho}(t)\ c^\dagger_{[z]+m}c_{[z]}\right]e^{ipm}\ .
\end{align}
As within $|A|$ our state is by assumption approximately homogeneous
and has a finite correlation length this corresponds to a
position-dependent density of states in momentum space, as advertised.

\item[\lbrack W4.\rbrack] 
The Heisenberg equations of motion for the Wigner function are linear
  \be
  \partial_t \hat{W}(z,p;t)+\epsilon'(p)\left[\hat{W}(z+\frac{1}{2},p;t)
-\hat{W}(z-\frac{1}{2},p;t)\right]=0.
\label{eomW}
  \ee
Given some initial values $W(z,p;0)$ the solution to \fr{eomW} is
obtain by Fourier methods
\be
W(z,p;t)=\sum_{z'}W(z',p;0)\int_{-2\pi}^{2\pi}\frac{dQ}{4\pi}e^{iQ(z-z')-2it\epsilon'(p)\sin(Q/2)}\ .
\ee

\end{itemize}
\subsubsection{Euler scaling limit}
Let us now consider the Euler scaling limit
\be
z=\Lambda\xi\ ,\quad t=\Lambda\tau\ ,\ \Lambda\to\infty\ , \xi,\tau
\text{ fixed}\ ,
\label{scalinglimit}
\ee
and assume that
\be
\lim_{\Lambda\to\infty}W(\Lambda\xi,p;\Lambda\tau)=\rho_{\xi,\tau}(p)
\ee
exists. Then we have from \fr{eomW} 
\be
\boxit{\partial_\tau
  \rho_{\xi,\tau}(p)+\epsilon'(p)\partial_\xi\rho_{\xi,\tau}(p)=0\ .}
\label{GHDFF2}
\ee
This has the same form as the evolution equation \fr{GHDFF}, and
by virtue of [W3] $\rho_{\xi,\tau}(p)$ can indeed be thought of as a
``ray-dependent'' mode occupation function. We note that the solutions
to \fr{GHDFF2} have simple expressions in terms of the ``initial
conditions'' at $\tau=0$
\be
\rho_{\xi,\tau}(p)=\rho_{\xi-\epsilon'(p)\tau}(p)\ .
\ee
In order for the Euler scaling limit to be non-trivial (in the sense
that one obtains ray-dependent results) the initial momentum space
Green's function ${\rm  Tr}[\hat\rho(0)c^\dagger(p)c(q)]$ must exhibit
some singularities as \fr{GFs} can otherwise be determined by a simple
stationary-phase approximation with saddle points only along the real
axis. 

\subsubsection{Expectation values of local operators in free theories}
Let us consider
expectation values of local operators along rays in spacetime. The
basic building block is the single-particle Green's function
\be
G(j,\ell,t)={\rm Tr}\left[\hat\rho(t) c^\dagger_jc_\ell\right]=\int
\frac{dp_1dp_2}{(2\pi)^2}e^{-ip_1j+ip_2\ell}
\underbrace{{\rm Tr}\left[\hat\rho(0)
    c^\dagger(p_1,t)c(p_2,t)\right]}_{\frac{1}{2}\sum_z
  e^{i(p_1-p_2)z}W(z,\frac{p_1+p_2}{2};t)}\ .
\ee
Let us then go over to the scaling variables \fr{scalinglimit}
\be
\frac{j+\ell}{2}=\Lambda\xi\ ,\quad t=\Lambda\tau\ ,
\ee
and change integration variables to
\be
p_+=\frac{p_1+p_2}{2}\ ,\quad p_-=\Lambda(p_1-p_2)\ .
\ee
In the limit $\Lambda\to\infty$ we have
\begin{align}
  \lim_{\Lambda\to\infty}G(j,\ell,t)&=
\lim_{\Lambda\to\infty}\int_{-\pi}^{\pi}\frac{dp_+}{2\pi}e^{-ip_+(j-\ell)}\int_{-2\Lambda(\pi-|p_+|)}^{2\Lambda(\pi-|p_+|)}\frac{dp_-}{2\pi}
\frac{1}{2\Lambda}\sum_z
e^{-ip_-(\xi-\frac{z}{\Lambda})} W(z,p_+;\Lambda\tau).
\end{align}
We then turn the sum over $z$ into an integral
\be
\frac{1}{2\Lambda}\sum_z
e^{-ip_-(\xi-\frac{z}{\Lambda})}
W(z,p_+;\Lambda\tau)\approx\int_{-\infty}^\infty dy
e^{-ip_-(\xi-y)}\rho_{y,\tau}(p_+)\ ,
\ee
which gives the final result
\begin{align}
\lim_{\Lambda\to\infty}G(j,\ell;t)
&\approx\int \frac{dp_+}{2\pi}
e^{-ip_+(j-\ell)}\rho_{\xi,\tau}(p_+)\equiv g_{\xi,\tau}(j-\ell)\ .
\end{align}
This is precisely what our GHD ``phenomenology'' predicts:
Interpreting $\rho_{\xi,\tau}(p)$ as a mode occupation
function along the ray characterized by $\xi$ and $\tau$ we can
construct a corresponding macro-state $|\rho_{\xi,\tau}\rangle$
by a family of Fock states with momenta distributed according to
$\rho_{\xi,\tau}(p)$. GHD then predicts that
\be
\lim_{\Lambda\to\infty}G(j,\ell;t)=
\langle\rho_{\xi,\tau}|c^\dagger_{j-\ell}c_0|\rho_{\xi,\tau}\rangle
=\int \frac{dp}{2\pi} e^{-ip(j-\ell)}\ \rho_{\xi,\tau}(p)\ .
\ee
As shown below in the scaling limit the anomalous Green's function
vanishes. This is expected \cite{essler2016quench} because the time
evolution operator has a U(1) symmetry $c_j\rightarrow
c_je^{i\varphi}$ related to particle number conservation. The initial
density matrix may break this symmetry, resulting in a non-vanishing
anomalous Green's function $F(j,\ell;0)$. However, as the system
relaxes locally the U(1) symmetry gets restored, which is why
$F(j,\ell;t)$ must vanish at late times. 
This means that the expectation value of an arbitrary local operators
along a given ray can be calculated by Wick's theorem using the
regular Green's function only. For example, along the ray
$t=\Lambda\tau$, $j_\alpha=\Lambda\xi+{\cal O}(\Lambda^0)$ 
\be
{\rm Tr}\left[\hat\rho(t)
  c^\dagger_{j_1}c^\dagger_{j_2}c_{j_3}c_{j_4}\right]\approx
g_{\xi,\tau}(j_1-j_4)g_{\xi,\tau}(j_2-j_3)
-g_{\xi,\tau}(j_1-j_3)g_{\xi,\tau}(j_2-j_4)\ .
\label{4point}
\ee

\subsubsection{Anomalous Green's function}
We can see that in the scaling limit the anomalous Green's function
vanishes by considering
\begin{align}
W_A(z,p;t)&={\rm Tr}\left[\hat\rho(0) \hat{W}_A(z,p;t)\right]\ ,\nn
\hat{W}_A(z,p;t)&=\int \frac{dq}{2\pi}c\big(p+\frac{q}{2}\big)
  c\big(-p+\frac{q}{2}\big)e^{iqz-it[\epsilon(p+\frac{q}{2})+\epsilon(p-\frac{q}{2})]}\ ,\quad 2z\in\mathds{Z}.
\end{align}
We have
\begin{itemize}
\item{} The anomalous Green's function can be recovered from ${g}$ using
  \be
{\rm Tr}\left[\hat\rho(t) c(p_1)c(p_2)\right]=\frac{1}{2}\sum_{2z\in\mathds{Z}}
e^{-iz(p_1+p_2)}W_A(z,\frac{p_1-p_2}{2};t)\ .
\label{AGF}
\ee
\item{} The Heisenberg equations of motion give
  \be
i  \partial_t W_A(z,p;t)=-2J\cos(p)\left[W_A(z+\frac{1}{2},p;t)
+W_A(z-\frac{1}{2},p;t)\right]-2\mu W_A(z,p;t).
  \ee
\item{} Going over to scaling variables this becomes
  \be
  \frac{1}{\Lambda}\partial_\tau
W_A(\Lambda\xi,p;\Lambda\tau)\approx-2i\epsilon(p)W_A(\Lambda\xi,p;\Lambda\tau)\ ,
  \ee
which shows that $W_A(\Lambda\xi,p;\Lambda\tau)$ vanishes in the
scaling limit. Going beyond the scaling limit we see that
the anomalous Green's function in position space in fact vanishes as a
power law in $t$, i.e. slowly. This is important, as it imposes a
restriction on the applicability of GHD: it will only work at times
when the anomalous Green's function is small \footnote{For example
there will be corrections to \fr{4point} involving the anomalous Green's function.}.
\end{itemize}

\subsubsection{Application of GHD away from the scaling regime}
As we have seen above, in the Euler scaling limit GHD becomes
exact. In practice one of course would like to apply it away from the
scaling limit as well. To that end one writes the GHD equation in the
original variables
\be
\partial_t\rho_{x,t}(p)+\epsilon'(p)\partial_x\rho_{x,t}(p)=0\ ,
\ee
and then uses that the solution of this equation is of the form
\be
\rho_{x,t}(p)=\rho_{x-\epsilon'(p)t,0}(p)\ .
\ee
The ``initial'' value $\rho_{x,0}(p)$ is then approximated by the
actual initial conditions
\be
\rho_{x,0}(p)\approx W(x,p;0)\ .
\label{initialcond}
\ee
Finally, the single-particle Green's function is approximated as
\be
G(j,\ell;t)\approx\int\frac{dp}{2\pi}e^{ip(j-\ell)}\ \rho_{x,t}(p)\ .
\ee
Using Wick's theorem this allows us to obtain approximate results for
multi-particle Green's functions as well.

We will see below how well this works in an explicit example.

\subsubsection{Corrections to GHD}
For free fermions it is possible to obtain a systematic expansion
around the GHD limit \cite{Fagotti17higher,Fagotti20locally}. In order
to do so we carry out a gradient expansion of \fr{eomW} beyond the
leading order, i.e. 
\be
\partial_t W(x,p;t)+\epsilon'(p)
\left[\partial_xW(x,p;t)+\frac{1}{24}\partial_x^3W(x,p;t)+\dots\right]=0. 
\ee
Importantly, we also must account for the fact that away from the
scaling limit the anomalous Wigner function does not vanish.

\subsection{Example: Free expansion}
\begin{figure}[ht]
\begin{center}
\epsfxsize=0.4\textwidth
\epsfbox{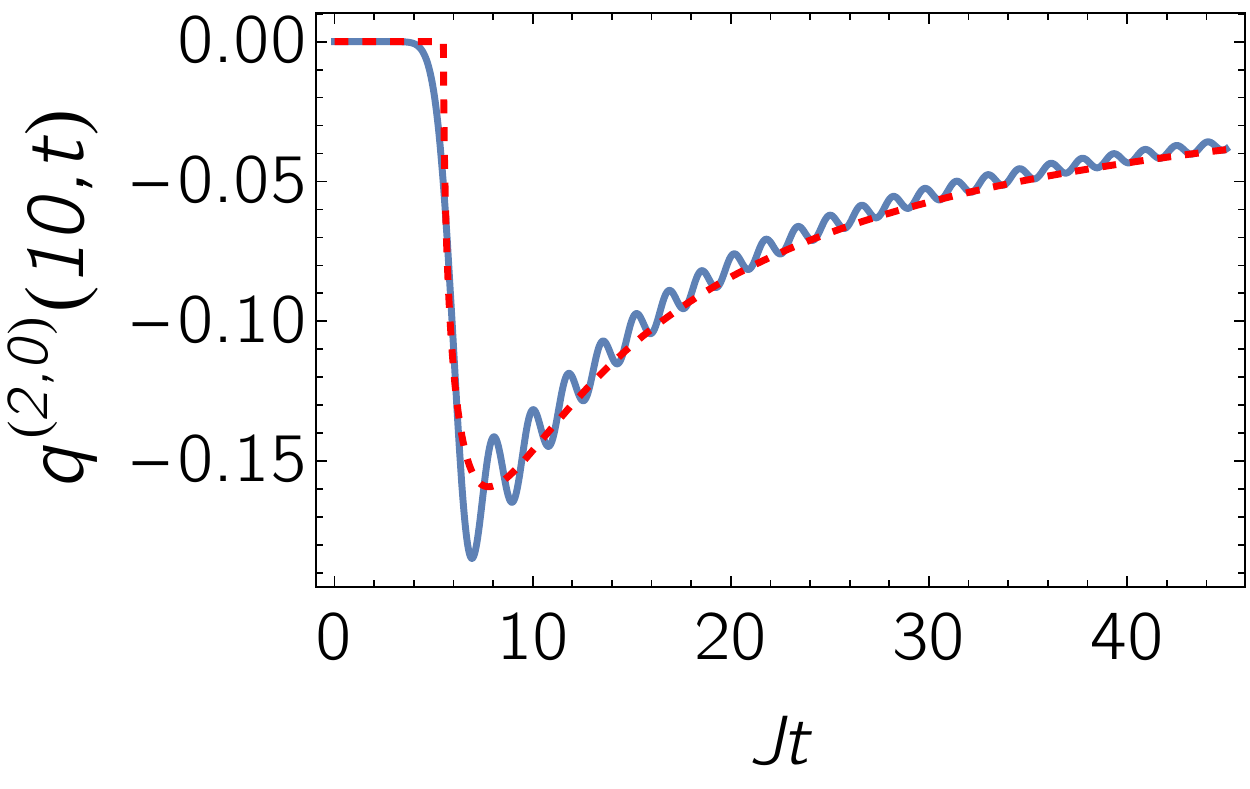}\qquad
\epsfxsize=0.4\textwidth
\epsfbox{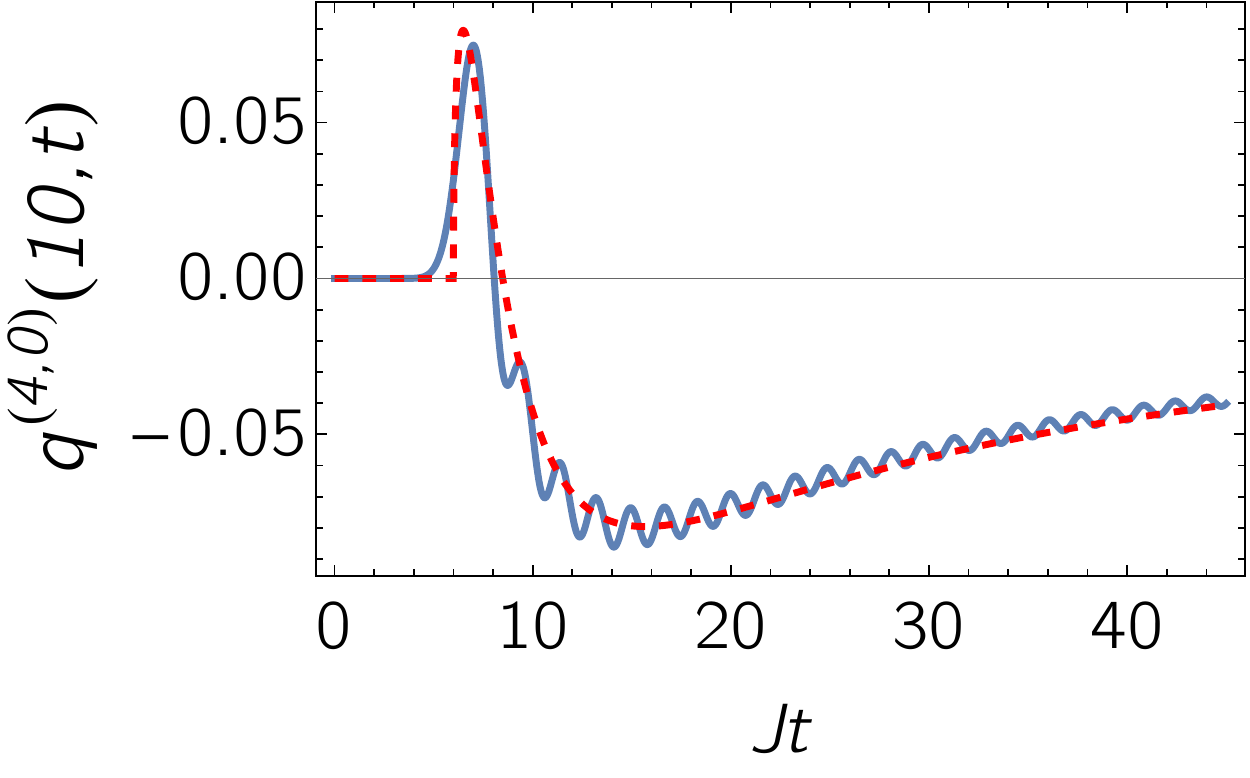}
\label{fig:q24}
\end{center}
\caption{Expectation values of the densities of the conserved
charges $q^{(2n,0)}(10,t)$ for $n=1,2$ as a function of time. The
solution to the GHD equation (dashed red line) is in fair agreement
with the exact result (solid blue line).}
\end{figure}

Let us now consider the explicit example, where our initial density
matrix correspond to a product state in position space where all sites
$j\leq 0$ are occupied
\be
|\Psi(0)\rangle=\prod_{j\leq 0}c^\dagger_j|0\rangle\ ,\quad
\hat\rho(0)=|\Psi(0)\rangle\langle\Psi(0)|\ .
\ee
We time-evolve this state with the tight-binding Hamiltonian
\fr{Hfree}. This problem is nice as it can be solved analytically
\cite{antal1999transport,viti2016inhomogeneous}. Using that
\be
{\rm  Tr}[\hat\rho(0)c^\dagger(p)c(q)]=\sum_{n\leq 0}e^{in(p-q)}\ ,
\ee
one finds 
\be
G(j,\ell;t)=\sum_{n=0}^\infty i^{\ell-j}J_{j+n}(2Jt)J_{\ell+n}(2Jt)\ ,
\ee
where $J_n(z)$ are Bessel functions. Using the addition theorem for
Bessel functions this can be simplified 
\be
G(j,j;t)=\frac{1}{2}\big[1-J_0^2(2Jt)\big]+
\begin{cases}
  -\sum_{m=1}^{j-1}J_m^2(2Jt) & \text{if }j>0\ ,\\
  \sum_{m=0}^{|j|}J_m^2(2Jt) & \text{if }j\leq 0\ .
\end{cases}
\ee
Similarly one has for $j>0$ and $n\geq 1$
\be
G(j,j+2n;t)=(-1)^n\Big\{-\frac{1}{2}\sum_{k=0}^{2n}(-1)^kJ_k(2Jt)J_{2n-k}(2Jt)
-\sum_{k=1}^{j-1}J_k(2Jt)J_{2n+k}(2Jt)\Big\}.
\ee
The application of GHD proceeds as follows. We first determine the
initial conditions for the Wigner function
\be
W(z,p;0)=\sum_{j\leq
  0}\frac{\sin\big(\pi(z-j)\big)}{\pi(z-j)}\approx\theta_H(-z)\ ,
\ee
where $\theta_H(x)$ is the Heaviside function.
Following the prescription \fr{initialcond} we then use this to fix
\be
\rho_{x,0}(p)=\theta(-x)\ ,
\ee
Here we have replaced the discrete lattice co-ordinate $z$ by a
continuous variable $x$. The rationale for doing this is that GHD is
expected to work on a length scale that is large compared to the
lattice spacing. The solution of the GHD equations is then 
\be
\rho_{x,t}(p)=\rho_{x-\epsilon'(p)t,0}(p)=\theta(-x+\epsilon'(p)t)\ .
\ee
Finally we use the relation between the Wigner function and the
Green's function that holds in the scaling limit to obtain an
approximate expression away from the scaling regime
\be
G(j,\ell;t)\approx\int
\frac{dp}{2\pi}e^{ip(j-\ell)}\ \rho_{\frac{j+\ell}{2},t}(p)=
\int
\frac{dp}{2\pi}e^{ip(j-\ell)}\ \theta\big(-\frac{j+\ell}{2}+\epsilon'(p)t\big).
\ee
\begin{figure}[ht]
\begin{center}
\epsfxsize=0.5\textwidth
\epsfbox{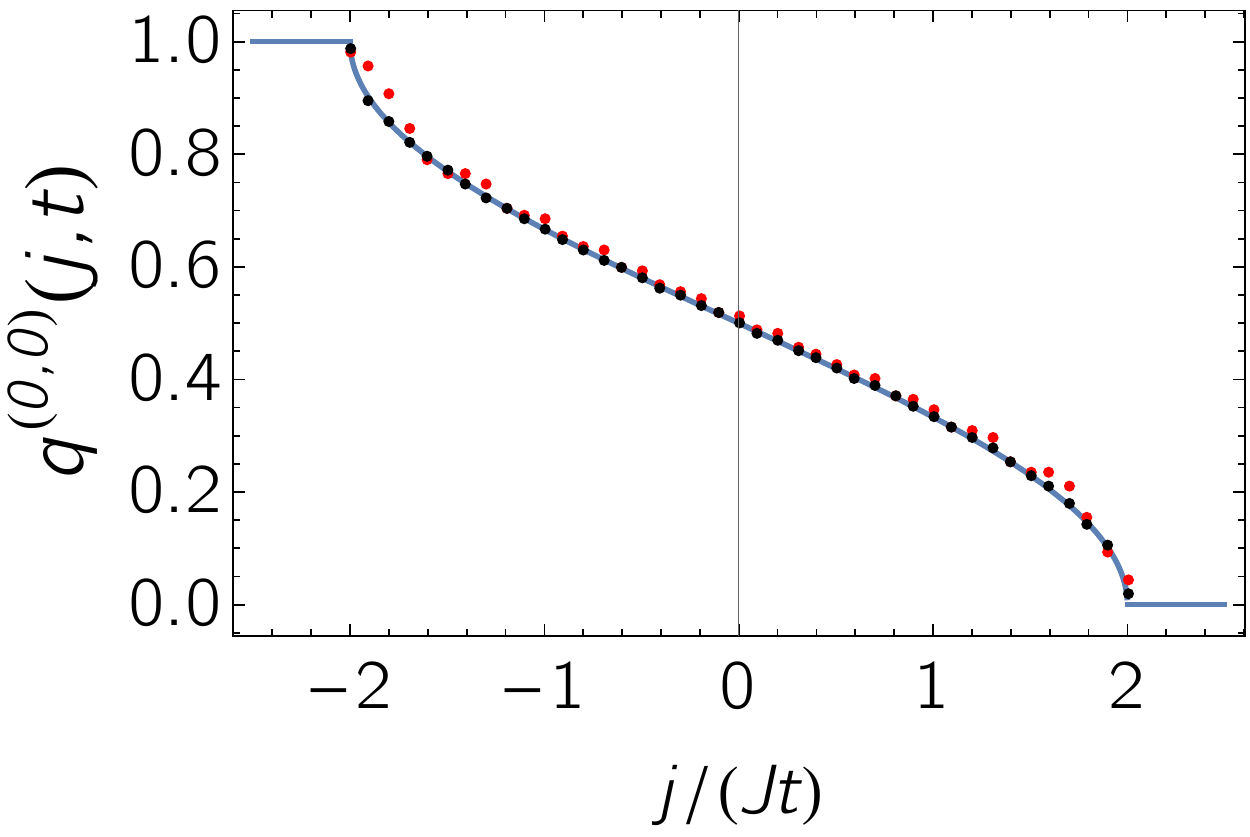}
\label{fig:profile}
\end{center}
\caption{Density profile along different ``rays'' $j/Jt$. Exact
  results at time $Jt=10$ (red dots) and $Jt=100$ (black dots) vs
Euler scaling limit (blue line).}
\end{figure}
We similarly can determine the expectation values of the densities of
the conserved charges \fr{charges}, e.g.
\begin{align}
q^{(2n,0)}(j+n,t)&\equiv\lim_{L\to\infty}\frac{1}{2}{\rm
  Tr}\big[\hat\rho(0)\big(c^\dagger_{j}c_{j+2n}+{\rm
    h.c.}\big)\big]\nn
&\approx\int
\frac{dp}{2\pi}\cos(2np) \theta\big(-j-n+\epsilon'(p)t\big)=
-\frac{\sin\big(2n\ {\rm arcsin}(\frac{j+n}{2Jt})\big)}{2\pi n}\theta_H\big(1-\frac{j+n}{2Jt}\big).
\end{align}
The result for the density profile in the Euler scaling limit is
\be
\lim_{\ontop{j,t\to\infty}{j/t \text{ fixed}}}{\rm Tr}\left[\hat{\rho}(t)c^\dagger_jc_j\right]
=\begin{cases}
1 & \text{if } j<-v_{\rm max}t\ ,\\
\frac{1}{\pi}{\rm arccos}\big(\frac{j}{2Jt}\big) & \text{if }
-v_{\rm max}t<j<v_{\rm max}t\ ,\\
0 & \text{if } j>v_{\rm max}t\ ,
\end{cases}
\ee
where $v_{\rm max}=2J$ is the maximal group velocity. In
Fig.~\ref{fig:profile} we compare this asymptotic result to the
density profile along rays $j/t=$fixed at finite times $Jt=10$ and
$Jt=100$. 
We observe that the GHD result provides a rather good approximation
already at times $Jt\sim10$. To get a more precise understanding how
the exact result approaches GHD in the scaling limit we show the
expectation values of $q^{(2n,0)}(10,t)$ for $n=1,2$ in Fig.~\ref{fig:q24}.
Initially these expectation values are zero because there are no
fermions at any of the sites $j>0$. At a time $v_{\rm max}t_n^*=10+n$ the
propagating front arrives at site $j=10+n$ and the expectation value
$q^{(2n,0)}(10,t)$ becomes sizeable. At later times the expectation
values approach the asymptotic GHD result in a power-law fashion. In
particular there is no characteristic time scale for local relaxation:
\emph{the system relaxes locally in a power-law fashion}. This is
expected to be a generic feature for integrable models.

\section{GHD in interacting integrable models}
For interacting integrable models we proceed along the same lines as
in the free fermion case:
\begin{enumerate}
\item{} Identify local conservation laws $Q^{(n)}=\sum_jQ^{(n)}_j$.
\item{} Construct simultaneous eigenstates of $Q^{(n)}$.
\item{} Construct macro-states in thermodynamic limit.
\item{} Work out stable excitations over macro-states and determine
  their group velocities.
\item{}  Apply a “local density approximation” to the continuity
  equations for the densities $Q^{(n)}_j$.
\end{enumerate}
We'll consider the implementation of this programme for the particular
example of the Lieb-Liniger model. The first four steps are discussed
in section (\ref{sec:LL}). In a homogeneous macro-state with density
$\rho(k)$ we have shown in section \fr{ssec:EVmacro} that in the
thermodynamic limit we have
\begin{align}
\langle\boldsymbol{ \rho}| Q^{(n)}(x)|\boldsymbol{ \rho}\rangle
&=\int\frac{d\lambda}{2\pi}q^{(n)}(\lambda)\ \rho(\lambda)\ .
\end{align}
The expectation values of the current densities are more difficult to
determine. A simple physically intuitive guess is 
\begin{align}
\langle\boldsymbol{ \rho}| J^{(n)}(x)|\boldsymbol{ \rho}\rangle
&=\int\frac{d\lambda}{2\pi}v_{\boldsymbol{\rho}}(\lambda)\ q^{(n)}(\lambda) \rho(\lambda)\ ,
\label{currexp}
\end{align}
where $v_{\boldsymbol{\rho}}(\lambda)$ is the group velocity of the stable
particle excitations with momentum $p$ over the macro-state with density
$\rho(\lambda)$, \emph{cf.} eqn \fr{vrho} in section
\ref{ssec:stable}. The main difference to the case of free fermions is
that this velocity now depends on the macro-state under
consideration. It turns out that the guess \fr{currexp} is indeed
correct \cite{pozsgay2020algebraic,borsi2021current}.
Now we proceed as in the case of free theories and first consider the
case where we prepare our system in a density matrix $\hat\rho(0)$
that locally corresponds to a homogeneous macro-state but varies
slowly in space. Locally it will ``look like'' a macro-state with a
position-dependent density $\rho_{x,0}(k)$, so that
\begin{align}
{\rm Tr}\left[\hat\rho(0)\ Q^{(n)}(x)\right]
&\approx\int\frac{d\lambda}{2\pi}q^{(n)}(\lambda) \rho_{x,0}(\lambda)\ ,\nn
{\rm Tr}\left[\hat\rho(0)\ J^{(n)}(x)\right]
&\approx\int\frac{d\lambda}{2\pi}v_{\boldsymbol{\rho}}(\lambda)\ q^{(n)}(\lambda) \rho_{x,0}(\lambda)\ .
\end{align}
Next we consider the situation at sufficiently late times after our
quantum quench, where we assume that our system has relaxed locally and
varies very slowly in space. Then we again can employ a description in
terms of an appropriately chosen macro-state, which however now will
depend on $x$ and $t$ 
\begin{align}
{\rm Tr}\left[\hat\rho(0)\ Q^{(n)}(x,t)\right]&=
{\rm Tr}\left[\hat\rho(t)\ Q^{(n)}(x)\right]
\approx\int\frac{d\lambda}{2\pi}q^{(n)}(\lambda) \rho_{x,t}(\lambda)\ ,\nn
{\rm Tr}\left[\hat\rho(0)\ J^{(n)}(x,t)\right]&=
{\rm Tr}\left[\hat\rho(t)\ J^{(n)}(x)\right]
\approx\int\frac{d\lambda}{2\pi}\ q^{(n)}(\lambda) v_{\boldsymbol{\rho}_{x,t}}(\lambda)\rho_{x,t}(\lambda)\ .
\label{QJint}
\end{align}
The evolution equation for $\rho_{x,t}(k)$ then follows from
combining \fr{QJint} with the continuity equations \fr{conteqint}
relating the expectation values of the charge and current densities,
which gives
\be
\int\frac{d\lambda}{2\pi}q^{(n)}(\lambda)\left[
\partial_t\rho_{x,t}(\lambda)+\partial_x
\bigg(v_{\boldsymbol{\rho}_{x,t}}(\lambda)\rho_{x,t}(\lambda)\bigg)
\right]=0\ ,\quad n=0,1,2,\dots
\ee
As the $q^{(n)}(\lambda)$ form a basis of functions on the real line
these imply the following form for the GHD equations in the
Lieb-Liniger model 
\be
\boxit{
\partial_t\rho_{x,t}(\lambda)+\partial_x
\bigg(v_{\boldsymbol{\rho}_{x,t}}(\lambda)
\ \rho_{x,t}(\lambda)\bigg)=0\ .}
\label{GHDFFint}
\ee
We stress that the group velocities
$v_{\boldsymbol{\rho}_{x,t}}(\lambda)$ depend on $\rho_{x,t}(\lambda)$ in
a non-linear way. As we have observed before, the group velocities
depend on the macro-state only through the occupation function
$\vartheta_{x,t}(\lambda)$ \fr{chirho}. Moreover, for a given root
density we can determine the corresponding occupation function from
\fr{TLBAE}, and \emph{vice versa}. Using this relation between
$\rho_{x,t}$ and $\vartheta_{x,t}$ we can therefore obtain an
equivalent formulation of GHD in terms of the latter
\be
\boxit{
\partial_t\vartheta_{x,t}(\lambda)+
v_{\vartheta_{x,t}}(\lambda)
\ \partial_x\vartheta_{x,t}(\lambda)=0\ .}
\label{GHDFFint2}
\ee
\subsection{Integrating the GHD equations}
In order to use GHD one requires an initial value for the occupation function
$\vartheta_{x,0}(\lambda)$. In practice this imposes serious
restrictions on the situations that can be handled. This is easy to
understand: from a physical perspective the occupation function
encodes the expectation values of the densities $Q^{(n)}(x)$ of the
infinite number of conserved charges. These are known in thermal
equilibrium, and using a local density approximation in presence of a
potential $V(x)$ can be determined in a broad class of inhomogeneous
thermal equilibrium states. Certain other situations can be handled by
the method introduced in Ref.~\onlinecite{fagotti2013stationary} for
homogeneous quantum quenches with lowly entangled initial states, but
the problem of determining $\vartheta_{x,0}(\lambda)$ given a general
initial state remains unsolved.

Given an initial value $\vartheta_{x,0}(\lambda)$ the GHD equation
\fr{GHDFFint} needs to be integrated using that the group velocities
are given in terms of $\vartheta_{x,t}(\lambda)$ through \fr{vrho}. An
open source Matlab code for carrying out this programme was introduced
in Ref.~\onlinecite{moller2020introducing}, see
https://integrablefluid.github.io/iFluidDocumentation/

Once $\vartheta_{x,t}(\lambda)$ and $\rho_{x,t}(\lambda)$ have been
obtained numerically they can be used to determine physical
properties. The simplest ones are the expectation values of the
charge and current densities \fr{QJint}. Extensions to certain other
operators are also possible \cite{doyon2018exact}.

\subsection{Extensions of GHD}
GHD has been extended in many interesting directions, and we
refer the interested reader to the reviews collected in the recent
special issue Ref.~\onlinecite{bastianello2021introduction}. One
such extension that is very important for applications of GHD to
cold-atom experiments is to approximately take into account the
effects of an external potential $V(x)$ in the Lieb-Liniger
model. This modifies the Hamiltonian to
\be
H=\int dx
\ \Phi^\dagger(x)\left[-\frac{\hbar^2}{2m}\partial_x^2+V(x)\right]\Phi(x)+c\int
dx \big(\Phi^\dagger(x)\big)^2\big(\Phi(x)\big)^2\ ,
\label{HV}
\ee
and importantly breaks integrability. Hence GHD no longer provides an
exact description in the Euler scaling limit. However, as long as
$V(x)$ can be considered as a small perturbation, it makes sense to
consider how it modifies the GHD equations. The answer was derived in
Ref.~\onlinecite{doyon2017anote} 
\be
\partial_t\rho_{x,t}(\lambda)+\partial_x
\big(v_{\boldsymbol{\rho}_{x,t}}(\lambda)
\ \rho_{x,t}(\lambda)\big)=\big(\partial_xV(x)\big)\partial_\lambda\rho_{x,t}(\lambda).
\label{GHDV}
\ee
The obvious question is on which time and length scales this evolution
equation provides an accurate description of the non-integrable
dynamics induced by \fr{HV}. The answer is not known, but on physical
grounds one would expect that if the initial state of the system is
only weakly inhomogeneous and the potential $V(x)$ is weak as well, \fr{GHDV}
may provide a good approximation at \emph{sufficiently short
  times}. This is because eventually the fact that $V(x)$ breaks
integrability will be felt and a description based on continuity
equations for conserved quantities must fail. On the other hand, as we
have seen above, in general GHD is expected to work only at
\emph{sufficiently late times} unless the initial state is judiciously
chosen \cite{bulchandani2017solvable}. It is important to keep these
considerations in mind when applying \fr{GHDV}.

\section{Applications of GHD to cold-atom experiments}
\begin{figure}[ht]
\begin{center}
\epsfxsize=0.46\textwidth
\epsfbox{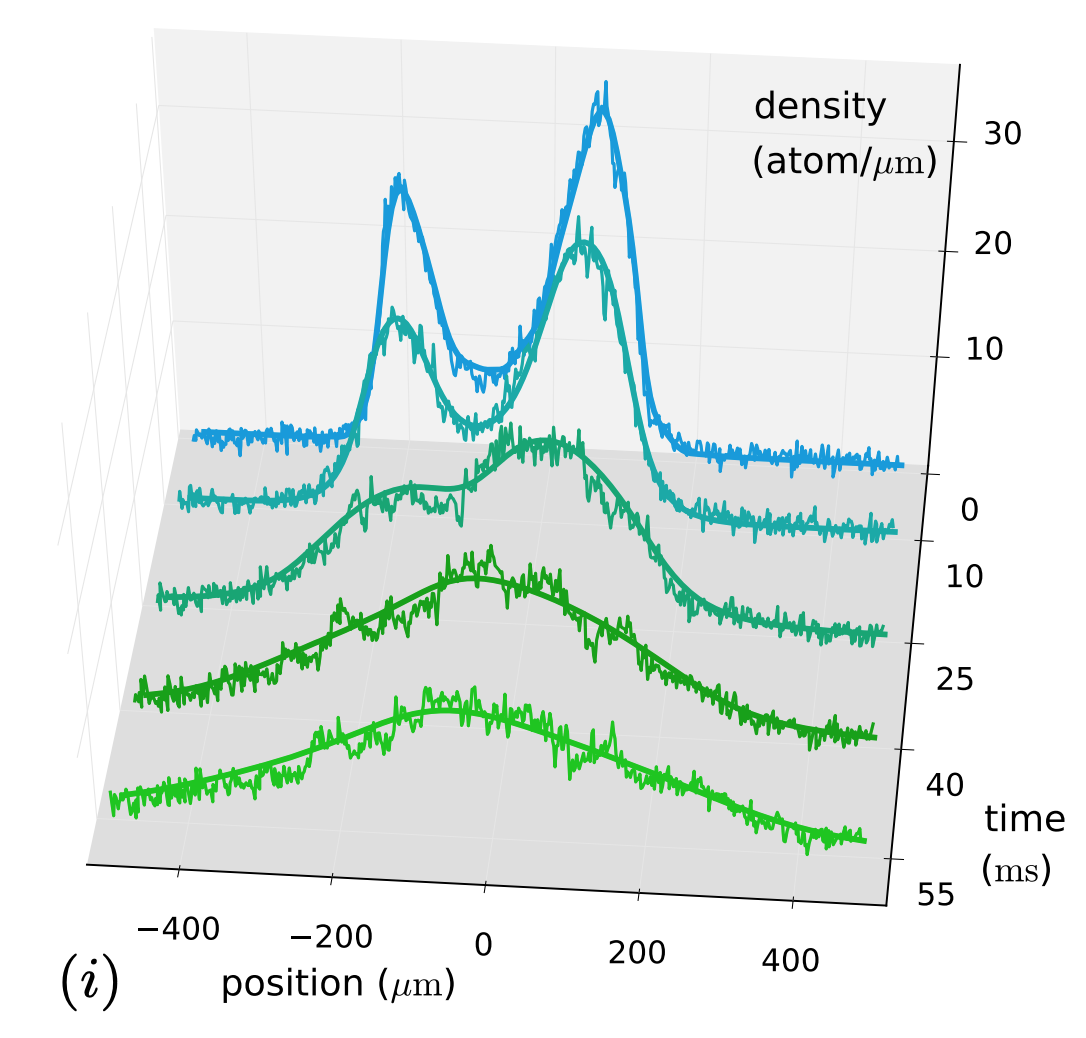}\qquad
\label{fig:exp}
\end{center}
\caption{Experimental results \emph{vs}  GHD prediction for atom
density during expansion from an initial thermal state in a
double-well potential (from
Ref.~\onlinecite{schemmer2019generalized}). The 
agreement is seen to be very good.}
\end{figure}

GHD and its extensions has been applied to the description of
cold-atom experiments that are described by the Lieb-Liniger model
with confining potential \fr{HV}, both in the weakly 
\cite{schemmer2019generalized} and in the strongly
\cite{malvania2021generalized} interacting regimes. A comprehensive
review of these works is presented in the recent review
\cite{bouchoule2022generalized}. In Fig.~\ref{fig:exp} we show some of
the results obtained in Ref.~\onlinecite{schemmer2019generalized}: a
gas of very weakly interacting ${}^{87}Rb$ atoms is prepared in a
thermal equilibrium state in a one-dimensional double-well potential,
i.e. like in the example discussed in section \ref{ssec:QQ} the
transverse confinement is very steep, but the potential along the
$y$-direction in Fig.~\ref{fig:split}(a) is a double-well rather than
harmonic. At time $t=0$ the confining potential along the
$y$-direction is changed to a shallow harmonic potential as in
Fig.~\ref{fig:split}(a), and the two clouds of gases start expanding
along the $y$-direction.
The density is then measured as a function of time. The main result,
and message of this section, is that the GHD prediction for this
expansion is found to be in very good agreement with experiment.

\acknowledgments
This work was supported in part by the EPSRC under grant
EP/S020527/1. I am grateful to Bruno Bertini and Etienne Granet for
very helpful discussions on the contents of these notes.

\bibliography{./bibliography}
\end{document}